\documentclass[12pt]{article}
\usepackage{dsfont}
\usepackage{makeidx}
\usepackage[utf8]{inputenc} 
\usepackage{times}
\usepackage{bm}
\usepackage{comment}
\usepackage{amssymb}
\usepackage{amsmath}
\usepackage{amsfonts}
\usepackage{pdflscape}
\usepackage{listings}
\numberwithin{equation}{section}
\usepackage{color}
\usepackage{graphicx}
\usepackage{rotating}
\usepackage{pdflscape}
\usepackage{epstopdf}
\usepackage[round,authoryear,comma]{natbib}
\usepackage{natbib,hyperref}
\usepackage{booktabs}
\usepackage{actuarialangle}
\usepackage{actuarialsymbol}
\usepackage{setspace}
\usepackage{physics}
\usepackage{pgfplots}
\usepackage{xcolor}
\usepackage{multirow}
\usepackage{caption}
\usepackage[lofdepth,lotdepth]{subfig}
\hypersetup{
	colorlinks   = true, 
	urlcolor     = blue, 
	linkcolor    = blue, 
	citecolor   = blue 
}

\providecommand{\U}[1]{\protect\rule{.1in}{.1in}}
\newtheorem{theorem}{Theorem}[section]

\newtheorem{remark}[theorem]{Remark}

\parskip=\medskipamount

\newcommand*{\PoissonDist}{\mathsf{Poisson}}

\usepackage{authblk}

\begin{document}

\title{Bayesian model averaging for mortality forecasting using leave-future-out validation\footnote{Contacts: Karim Barigou \url{karim.barigou@univ-lyon1.fr} (corresponding author), Pierre-Olivier Goffard \url{pierre-olivier.goffard@univ-lyon1.fr}, Stéphane Loisel \url{stephane.loisel@univ-lyon1.fr} and Yahia Salhi \url{yahia.salhi@univ-lyon1.fr}.}}
\date{Version: \today }

\author[1]{Karim Barigou}
\author[1]{Pierre-Olivier Goffard}
\author[1]{Stéphane Loisel}
\author[1]{Yahia Salhi}
\affil[1]{\small Univ Lyon, Universit\'e Claude Bernard Lyon 1, Laboratoire de Sciences Actuarielle et Financière, Institut de Science Financi\`ere et d'Assurances (50 Avenue Tony Garnier, F-69007 Lyon, France)}

\maketitle

\begin{abstract}
Predicting the evolution of mortality rates plays a central role for life insurance and pension funds.
Various stochastic frameworks have been developed to model mortality patterns taking into account the main stylized facts  driving these patterns. However, relying on the prediction of one specific model can be too restrictive and lead to some well documented drawbacks including model misspecification, parameter uncertainty and overfitting. To address these issues we first consider mortality modelling in a Bayesian Negative-Binomial framework to account for overdispersion and the uncertainty about the parameter estimates in a natural and coherent way. Model averaging techniques are then considered as a response to model misspecifications. In this paper, we propose two methods based on leave-future-out validation which are compared to the standard Bayesian model averaging (BMA) based on marginal likelihood. An intensive numerical study is carried out over a large range of simulation setups to compare the performances of the proposed methodologies. An illustration is then proposed on real-life mortality datasets which includes a sensitivity analysis to a Covid-type scenario. Overall, we found that both methods based on out-of-sample criterion outperform the standard BMA approach in terms of prediction performance and robustness.
\vspace{5mm}


\end{abstract}

\pagebreak
\section{Introduction}
\label{sec:introduction} 

Apart from short epidemic shocks, most developed countries face unprecedented improvements in longevity that contribute to the aging of the population. As a consequence, pension funds, social security systems and life insurers face longevity risk, namely the risk that policyholders live longer than expected. These concerns have led to an extensive development of stochastic mortality models in the actuarial, demographic and statistical literature. The selection of a specific model is naturally subject to model risk, that is the risk of picking the wrong model. This paper considers a full Bayesian model averaging approach to mitigate this risk while taking into account the uncertainty in the value of the parameters due to the potential lack of fit of the mortality models to the data.

A major part of the literature on stochastic mortality modelling has developed from the seminal work of \cite{lee1992modeling}. It introduced a factor-based framework on which the mortality surface (on the logarithmic scale) is decomposed into the sum of an age-specific term representing the average mortality rate per age and a bilinear term including a single time-varying index, which represent the mortality trend and an age-specific component that characterizes the sensitivity to this trend at different ages. Several extensions were proposed in the literature. For example, \cite{renshaw2006cohort} proposed an extension of the Lee-Carter model with a cohort effect and \cite{cairns2006two} proposed a two-factor model for pensioners mortality often abbreviated as CBD. The CBD model was then extended by incorporating combinations
of a quadratic age term and a cohort effect term in \cite{cairns2009quantitative}. \cite{plat2009stochastic} combined the features of existing models to come up with a model that covers the entire age range and takes into account cohort effects. For an overview of existing models, we refer to \cite{hunt01}. Mortality forecasts are usually obtained in a frequentist two-step procedure. In a first step, estimates of the parameters are obtained by Singular Value Decomposition or Maximum Likelihood Estimation, noticing that standard mortality models can be expressed as a generalized non-linear or linear model, see \cite{currie2016fitting}. In a second step, parameters are projected using time-series techniques. 

In this paper, we consider mortality modelling in a Bayesian framework. When compared to the classical framework, the Bayesian approach offers two notable advantages. First, the estimation and forecasting steps go hand in hand, which leads to more consistent estimates, see \cite{cairns2011bayesian} and \cite{wong2018bayesian} among others. Second, it better accounts for the different sources of uncertainty in a natural and coherent way. Within the literature on Bayesian mortality modeling, \cite{czado2005bayesian} proposed a fully integrated Bayesian  approach tailored to the  Poisson Lee-Carter (LC) model. It was extended to the multi-population setting in \cite{antonio2015bayesian}. \cite{pedroza2006bayesian} performed mortality forecasting using a Bayesian state-space model using Kalman filters, that handle missing data. \cite{kogure2010bayesian} presented a Bayesian approach to pricing longevity risk under the LC framework. Finally, \cite{venter2018parsimonious} considered Bayesian shrinkage to obtain a parsimonious parameterization of mortality models.

To account for model uncertainty, we consider model averaging. Compared to using the predictions of one specific model, combining the forecasts of various models is more robust toward model mis-specification and is more likely to produce reliable point and interval forecasts. There are two standard approaches to model averaging: a frequentist approach based on the Akaike Information Criterion (AIC) by \cite{buckland1997model} and a Bayesian approach known as Bayesian model averaging, see \cite{hoeting1999bayesian}, relying on the Bayes factor, see \cite{Kass1995}. While both approaches received much attention in several areas such as ecology (\cite{cade2015model}) or finance (\cite{koop2012forecasting}), there are only few papers in the context of demography and actuarial science. \cite{shang2012point} combined mortality forecasts based on two weighting schemes, the first is based on out-of-sample forecast accuracy and the other relies on in-sample goodness-of-fit. \textcolor{black}{Instead of choosing the optimal weights, \cite{shang2018model} considered selecting a subset of superior models before equally averaging forecasts from these selected models.} In the Bayesian setting, we only found \cite{benchimol2018mortality} who applied Bayesian Model Averaging (BMA) to combine four popular mortality models via their posterior probability. However, they did not show the mathematical details nor did they compare the BMA with the single-model forecasts.

In this paper, we propose a full Bayesian approach for mortality forecasting. We first sample from the posterior distribution of the mortality model parameters using Markov Chain Monte Carlo (MCMC) techniques. We then derive weights for each mortality model. The standard method for calculating the Bayesian model weights uses a marginal likelihood approximation. The latter characterizes the suitability of the model to the data used to train this very model. We therefore introduce two alternative model averaging methods based on the forecast accuracy measured on a validation data set (different from the training one). The validation set is made of the most recent years, hence the name \textit{leave-future-out} validation. We refer to these method as \textit{stacking} and \textit{pseudo-BMA} because they follow from an adaptation of the model averaging strategies described in the work of \cite{yao2018using} based on leave-one-out validation. We show that stacking and pseudo-BMA outperform the standard averaging approach in terms of forecasting accuracy when applied to real as well as simulated mortality data. 
To the best of our knowledge, this is the first time that a Bayesian model averaging approach based on out-of-sample performance is considered for mortality forecasting.

The remainder of the paper is organized as follows. In \autoref{sec:BMMC}, we introduce the Bayesian mortality modeling framework which accommodates a wide range of well-known mortality models. In \autoref{sec:BMMA}, we discuss model aggregation strategies, starting with the standard method before moving on to the alternative methods designed to make predictions. In \autoref{sec:sim}, an intensive numerical study is carried out accross a large range of simulation setups to provide a fair comparison of the proposed methodologies. \autoref{sec:real_data} compares the prediction performance of the model averaging methodologies on real-life mortality datasets. \autoref{sec:covid} investigates the impact of a COVID-type effect on the mortality rate projections and \autoref{sec:conclusion} provides some concluding remarks and perspectives for future research work. 

\section{Bayesian Mortality Modeling}\label{sec:BMMC}
When studying human mortality, the data at hand consist of death counts \(d_{x,t}\) and central exposures \(e_{x,t}\), where \(x=x_1, x_2,\ldots, x_A\) and 
\(t = t_1,t_2,\ldots,t_N\) represent a set of $A$ age groups and $N$ calendar years respectively. We denote by \(\mu_{x,t}\) the force of mortality at age \(x\) and calendar year \(t\). A stochastic mortality model commonly relies on two assumptions: 
\begin{enumerate}
	\item \textcolor{black}{The number of deaths is modelled by a counting random variable \(D_{x,t}\) following either a Poisson, binomial, or negative binomial distribution.}
	\item The force of mortality has a log or logit link to the age and calendar year variables.
\end{enumerate}

\subsection{Negative-Binomial model}
The data provided to mortality models are generally at the country level. Empirical studies have shown that life expectancy depends on socioeconomic status, individual income, education, marital status, among other factors. This heterogeneity within a given population tends to increase the variability of the underlying death counts, leading to overdispersion. To tackle this issue, we consider a classic extension of the Poisson distribution, namely a gamma mixture of Poisson distributions, which assumes that 
\begin{align}
D_{x,t}\mid \mu_{x,t} &\stackrel{\text { ind }}{\sim} \PoissonDist(\mu_{x,t} e_{x,t})\label{eq:death_count_dist}\\
\log \mu_{x,t} &= \alpha_x +\sum_{i=1}^{p}\beta_x^{(i)}\kappa_{t}^{(i)}+ \beta_{x}^{(0)}\gamma_{t-x} + \log \nu_{x,t}\\
\nu_{x,t}\mid \phi &\stackrel{\text { ind }}{\sim} \operatorname{Gamma}(\phi, \phi),
\end{align}
where the average mortality rate within each age group relates to the \(\alpha_x\) coefficient, while age specific patterns of mortality improvement over time are captured through the \(\beta_x^{(i)}\) and \(\kappa_t^{(i)}\) for \(i=1,\ldots, p\). The model can accomodate for an age-specific cohort effect with the product of \(\beta_{x}^{(0)}\) by \(\gamma_{t-x}\) while overdispersion relates to the parameter $ \phi$. The expectation and variance of this model are given by 
\begin{align}
\mathbb{E}\left[D_{x,t}\right]&=e_{x,t}\exp \left(\alpha_x +\sum_{i=1}^{p}\beta_x^{(i)}\kappa_{t}^{(i)}+ \beta_{x}^{(0)}\gamma_{t-x}\right)\\
\operatorname{Var}\left[D_{x,t}\right]&=\mathbb{E}\left[D_{x,t}\right] \times\left[1+\frac{\mathbb{E}\left[D_{x,t}\right]}{\phi}\right]>\mathbb{E}\left[D_{x,t}\right].\label{varnb}
\end{align}
This model has the same mean as the standard Poisson model but possesses a larger variance which depends on the value of $\phi$. When $\phi \rightarrow \infty$, we recover the standard Poisson model. An important feature of this model is its equivalence to a Negative-Binomial (NB) model, in the sense that
\begin{equation*}
D_{x,t}\mid \alpha_x,\beta_{x},\kappa_{t},\gamma_{t-x},\phi \sim \operatorname{Neg}-\operatorname{Bin}\left(e_{x,t}\exp \left(\alpha_x +\sum_{i=1}^{p}\beta_x^{(i)}\kappa_{t}^{(i)}+ \beta_{x}^{(0)}\gamma_{t-x}\right),\phi\right).
\end{equation*}
The NB model was considered in a frequentist framework by \cite{delwarde2007negative} and in a Bayesian setting by \cite{wong2018bayesian}. We remark that \cite{wong2018bayesian} compared the NB model with a Poisson model with normal random error $\nu_{x,t}$, and found that both specifications provide similar fits. 

Under the NB assumption, the full likelihood of the death records is given by 
\begin{equation}\label{likelihood}
l(y \mid \boldsymbol{\alpha}, \boldsymbol{\beta}, \boldsymbol{\kappa},\boldsymbol{\gamma}, \phi)= \prod_{x, t} \left\{
\frac{\Gamma\left(d_{x t}+\phi\right)}{\Gamma(\phi) \Gamma\left(d_{x t}+1\right)} \left[\frac{e_{x t} \exp \left(\eta_{x,t}\right)}{e_{x t} \exp \left(\eta_{x,t}\right)+\phi}\right]^{d_{x t}} 
\left[\frac{\phi}{e_{x t} \exp \left(\eta_{x,t}\right)+\phi}\right]^{\phi}
\right\},
\end{equation}
with \begin{equation}\label{etaequation}
	\eta_{x,t}=\alpha_x +\sum_{i=1}^{p}\beta_x^{(i)}\kappa_{t}^{(i)}+ \beta_{x}^{(0)}\gamma_{t-x}.
\end{equation}

\noindent In this section we are concerned with finding the parameters 
\[
\theta = (\alpha_x,\beta_x^{(0)},\ldots, \beta_{x}^{(p)}, \kappa_t^{(1)},\ldots, \kappa_{x}^{(p)},\gamma_{t-x},\phi),
\]
in the set of possible parameters $\Theta$,  that best explains our data $y = (d_{x,t},e_{x,t})$, for \(x=x_1,\ldots, x_A\) and \(t =t_1,\ldots, t_N\). 

\subsection{Bayesian analysis}

Bayesian inference is based on the idea of updating our prior beliefs \(p(\theta)\) over \(\theta\) with the observed data at hand \(y\) to come up with posterior beliefs $p(\theta|y)$, see \cite{Gelman1995}. By Bayes' theorem, we can determine the posterior distribution of the parameters given the data as follows
\begin{equation}
p(\theta|y)=\frac{p(y|\theta)p(\theta)}{\int_{\Theta}p(y|\theta)p(\theta)},\label{eq:posterior_dist}
\end{equation}
which in turn allows us to build credible intervals as well as point estimates of the parameters by taking the mean or the mode of the posterior. The integral in the denominator of \eqref{eq:posterior_dist} is often analytically intractable due to the high dimension of the parameter space \(\Theta\). The usual workaround consists in sampling from the posterior distribution using a Markov Chain Monte Carlo (MCMC) simulation scheme: 
\begin{equation*}
\theta^{(1)},\theta^{(2)}, \ldots, \theta^{(M)} \sim p(\theta|y) \propto p(y|\theta)p(\theta).
\end{equation*}

In this paper, we consider five standard mortality models, each implemented in the Negative-Binomial setting with the likelihood \eqref{likelihood}. In \autoref{mortalitymodels}, we specify the predictor $\eta_{x,t}$ entering in the likelihood through \eqref{etaequation}. Hereafter, we discuss the prior distributions of the different parameters.

\begin{table}[h]
	\centering 
		\caption{Model structures considered in this paper.}\label{mortalitymodels}
	\begin{tabular}{ll}
		\toprule
		Mortality model & Predictor $\eta_{x,t} $ \\
		\midrule
		Lee-Carter (LC) & $\eta_{x,t}=  \alpha_x+\beta_x \kappa_t^{(1)}$ \\
		Renshaw-Haberman (RH) & $\eta_{x,t}=  \alpha_x+\beta_x \kappa_t^{(1)}+ \gamma_{t-x}$ \\
		Age-Period-Cohort (APC) & $\eta_{x,t}=  \alpha_x+ \kappa_t^{(1)}+\gamma_{t-x}$ \\
		Cairns-Blake-Dowd (CBD) & $\eta_{x,t}= \kappa_t^{(1)}+(x-\bar{x})\kappa_t^{(2)}$ \\
		M6 & $\eta_{x,t}= \kappa_t^{(1)}+(x-\bar{x})\kappa_t^{(2)}+ \gamma_{t-x}$\\
		\bottomrule
	\end{tabular}
\end{table}

\subsection{Prior distributions}

For the choice of the prior distributions, there are essentially two common approaches. The first one specifies diffuse or weakly informative priors such that the posterior inference is dominated by the likelihood of the data, see e.g. \cite{wong2018bayesian}. The second one specifies prior distributions which depend on hyperparameters which are estimated by an empirical frequentist approach, see e.g. \cite{czado2005bayesian} and \cite{kogure2010bayesian}. In this paper, we follow the first approach.  

\subsubsection{Prior distribution for $\alpha_x,\beta_{x}$ and $\phi$}

Similar to \cite{wong2018bayesian}, we assign independent normal priors on $\alpha_x$, i.e. 
\begin{equation*}
	\alpha_x \sim N(\alpha_0,\sigma^2_{\alpha}),
\end{equation*}
with $\alpha_0=0$ and $\sigma^2_{\alpha}=100$. Because of the constraint $\sum_x \beta_{x}=1$, we let the $\beta_{x}$'s be Dirichlet distributed with 
\begin{equation*}
	\beta_{x} \sim \text{Dirichlet}(1,\dots,1).
\end{equation*}
Since the model variance is measured by $1/\phi$, see Equation \eqref{varnb}, the parameter $\phi$ is actually a concentration parameter for which a standard prior assumption is the half-normal distribution,
\begin{equation*}
	\frac{1}{\phi} \sim \text{Half-Normal}(0,1),
\end{equation*}
see for instance \cite{gelman2006prior}. 

\subsubsection{Prior distributions for $\kappa_{t}$}

For the period indexes we follow the standard actuarial science practice (\cite{cairns2011bayesian}, \cite{cairns2006two}, \cite{haberman2011comparative}, \cite{lovasz2011analysis}) and assume that the period indexes follow a multivariate random walk with drift. That is,
\begin{equation}
\bm{\kappa}_{t}=\bm{c}+\bm{\kappa}_{t-1}+\boldsymbol{\epsilon}_{t}^{\kappa},\quad \bm{\kappa}_{t}=\left(\begin{array}{c}
\kappa_{t}^{(1)} \\
\kappa_{t}^{(2)}
\end{array}\right), \quad \boldsymbol{\epsilon}_{t}^{\kappa} \sim N\left(\mathbf{0}, \Sigma\right),\label{dynamicskappa}
\end{equation}
where $\bm{c}$ is a $2$-dimensional vector of trend parameters and $\Sigma$ is a $2\times 2$ variance-covariance matrix of the multivariate white noise $\boldsymbol{\epsilon}_{t}^{\kappa}$. For models with a single period effect like LC, RH and APC, the dimension of Equation \eqref{dynamicskappa} shrinks to one. For the sake of identifiability, we impose $\kappa_1=0$ similar to \cite{haberman2011comparative} and \cite{wong2018bayesian}. Under this constraint, the remaining $ \kappa_{t}$ quantify the mortality improvements relative to the first year while the first year log mortality rates are determined by the $\alpha_x$'s. To complete the model specifications on the $\bm{\kappa}_{t}$'s, we set independent normal priors over the regression coefficients $\bm{c}\sim N(0,10)$. The variance-covariance matrix of the error term is defined by 
\begin{equation*}
\boldsymbol{\Sigma}=\left(\begin{array}{cc}
\sigma_1^{2} & \rho_{\Sigma} \sigma_1 \sigma_2 \\
\rho_{\Sigma} \sigma_1 \sigma_{Y} & \sigma_2^{2}
\end{array}\right)
\end{equation*}
where the variance coefficients are independent exponentials $\sigma_1, \sigma_2 \sim Exp(0.1)$ and the correlation parameter is uniform $\rho_{\Sigma} \sim U\left[-1,1\right]$.

\subsubsection{Prior distributions for $\gamma_{c}$}

For the cohort effect, we consider a second order autoregressive process (AR(2)): 
\begin{equation}\label{dynamicsgamma}
\gamma_{c}=\psi_1 \gamma_{c-1}+\psi_2 \gamma_{c-2}+\epsilon^{\gamma}_{t},\quad \epsilon^{\gamma}_{t}\sim N(0,\sigma_{\gamma}),
\end{equation}
which is in line with previous study conducted by \cite{cairns2011mortality} and \cite{lovasz2011analysis}. Several model specifications such as $\text{AR}(1)$ or $\text{ARIMA}(1,1,0)$ can be seen as special cases of Equation \eqref{dynamicsgamma}. To ensure identifiability, the cohort component is constrained so that the first and last components are equal to 0: 
\begin{equation*}
\gamma_1=0, \quad \gamma_{C}=0.
\end{equation*}
For the RH model, we also impose that the sum of effects over the whole range of cohorts is zero:
\begin{equation*}
\sum_{i=1}^{C} \gamma_{i}=0,
\end{equation*}
where $C$ corresponds to the most recent cohort. These constraints ensure that $\gamma$ truly represents a cohort effect. Indeed, if the cohort effect presents a trend, this can be compensated by an adjustement to the age and period effects. We close the model specification by imposing some vague priors assumptions on the hyperparameters: 
\begin{equation*}
	\psi_1,\psi_2 \sim N(0,10),\quad \sigma_{\gamma}\sim Exp(0.1).
\end{equation*}

\begin{remark}
\textcolor{black}{It is well-known that the RH model may have convergence issues \citep{currie2016fitting,hunt2015robustness}. Following \cite{cairns2011bayesian}, we first started our analysis with a stationary AR(2) process with constraints on the first and last component but found convergence issues during our simulation study (in 80 simulations, around 20 calls were not convergent). Adding the sum-to-zero constraint by imposing $\gamma_2=-\sum_{i=3}^{C-1}\gamma_i$ solved the convergence problem.}
\end{remark}

\subsection{Hamiltonian Monte Carlo and Stan}
To produce samples from the posterior distribution, we have implemented our stochastic mortality models using a programming language called \textit{Stan}, see \cite{carpenter2017stan}. Stan performs an Hamiltonian Monte Carlo (HMC) sampling scheme through the No-U-TurnS (NUTS) algorithm. \textcolor{black}{Compared with the random-walk Metropolis algorithm, where a proposed value is not related to the target distribution, HMC proposes a value that uses the derivatives of the density function being sampled to generate efficient transitions spanning the posterior (see e.g. \cite{neal2011mcmc} for details). It uses an approximate Hamiltonian dynamics simulation based on numerical integration which is then corrected by performing a Metropolis acceptance step.} HMC enhances the sampling efficiency and robustness for models with complex posteriors compared to the widely used Metropolis-Hasting within Gibbs sampling scheme.\footnote{\cite{neal2011mcmc} analyzes the scaling benefit of HMC with dimensionality. \cite{JMLR:v15:hoffman14a} provide practical comparisons of \textbf{Stan}’s adaptive HMC algorithm with Gibbs, Metropolis, and standard HMC sample.} The NUTS algorithm, introduced by \cite{JMLR:v15:hoffman14a}, cope with the difficult choice of the tuning parameters and makes possible the incorporation of the HMC routine into inferencial engines such as Stan. The latter software is gaining popularity among Bayesian statistics practitionners and actuarial scientists, see for instance the work of \cite{Gao2019} and \cite{hilton2019projecting} where Stan is used for claim reserving and mortality modeling, respectively. 

\paragraph{Implementation in R} We have built our own R package \texttt{StanMoMo} which implements the mortality models of \autoref{mortalitymodels} under the Poisson and the Negative-Binomial setting. It can be downloaded from \url{https://CRAN.R-project.org/package=StanMoMo}. The package provides high-level R functions to perform Bayesian mortality inference, model selection and model averaging while using \texttt{Stan} and HMC sampling in the background. 

\paragraph{HMC sampling} For each model, four parallel chains are constructed, each of length $4000$. The first half of each chain is used as a warm-up round (during which \texttt{stan} tunes the algorithm to reflect the characteristics of the posterior) and discarded. Parallel chains are used to better assess the convergence toward the posterior distribution. \textcolor{black}{During our analysis, we carefully checked that there were no diverging transitions and we also followed the diagnostic measure $\widehat{R}$ that was advocated by \cite{vehtari2020rank}. We checked that $\widehat{R}<1.01$ as recommended by the authors, which indicates that all parameters have converged to an acceptable degree.} The remainder of all the chains are then gathered and used for inference.

\section{Bayesian mortality model averaging}\label{sec:BMMA}
Instead of choosing one model, model averaging stems from the idea that a combination of candidate models among a model list $\mathcal{M}=(M_1,\dots,M_K)$ may perform better than one single model. The standard Bayesian approach, called \textit{Bayesian model averaging} (BMA), consists in weighing each model by its posterior model evidence. This approach is discussed in \autoref{sec::BMA1} but should be avoided for mortality forecasting for several reasons. Among them, BMA is very sensitive to prior choices and tends to select only one model asymptotically. Moreover, like the Bayes Information Criterion (BIC), BMA measures how well the model fits the past but not how well the model predicts the future.

We propose two alternative model averaging approaches, called \textit{stacking} and \textit{Pseudo-BMA}, based on leave-future-out and inspired from the work of \cite{yao2018using}. These approaches, seemingly more suited for forecasting, are described in \autoref{sec::BMA2}. 

\subsection{Bayesian model averaging by marginal likelihoods}\label{sec::BMA1}
In the standard BMA approach, each model is weighted by its posterior probability \begin{equation}\label{weightsbma}
p\left(M_{k} \mid y \right)=\frac{p\left(y \mid M_{k}\right) p\left(M_{k}\right)}{\sum_{k=1}^{K} p\left(y \mid M_{k}\right) p\left(M_{k}\right)},
\end{equation}
where
\begin{equation}\label{ML}
p\left(y \mid M_{k}\right)=\int_{\Theta} p\left(y \mid \theta_{k}, M_{k}\right) p\left(\theta_{k} \mid M_{k}\right) d \theta_{k},
\end{equation}
for $k\in\{1,\ldots, K\}$, is called the Marginal Likelihood (ML). 
The posterior distribution for any quantity of interest $\Delta$ (e.g. mortality forecasts) is then given by 
$$
p(\Delta \mid y)=\sum_{k=1}^{K} p\left(\Delta \mid M_{k}, y \right) p\left(M_{k} \mid y \right).
$$ 
Since we typically assume equal prior model probabilities, i.e. $p\left(M_{k}\right)=\frac{1}{K}$, it remains to compute the MLs for each model. \textcolor{black}{To do so, we use an importance sampling technique known as bridge sampling. The underlying principle is briefly recalled hereafter Let
$$
p_i(\theta) = \frac{\eta_i(\theta)}{Z_i},\text{ }i\in\{1,2\}.
$$
be two probability distributions known up to a normalizing constant $Z_i,i\in\{1,2\}$ and let $\theta\mapsto h(\theta)$ be a ``bridge" function. The normalizing constant  ratio $Z_1/Z_2$ may be written as
$$
r=\frac{Z_1}{Z_2} = \frac{\mathbb{E}_{ p_2}(\eta_1\cdot h)}{\mathbb{E}_{ p_1}(\eta_2\cdot h)},
$$ 
where $\mathbb{E}_{ p_i}$ stands for the expectation under $p_i(\theta)\text{, }i \in\{1,2\}$, and be approximated by 
\begin{equation}\label{eq:bridge_sampling_estimator_ratio}
\frac{Z_1}{Z_2}\approx\frac{\sum_{j = 1}^N\eta_1\left(\theta^{(2)}_j\right)h\left(\theta^{(2)}_j\right)}{\sum_{j = 1}^N\eta_2\left(\theta^{(1)}_j\right)h\left(\theta^{(1)}_j\right)},
\end{equation}
where $\theta^{(i)}_1,\ldots \theta^{(i)}_N\sim p_i(\theta),\text{ } i\in\{1,2\}$. The optimal bridge function from the quadratic error point of view is given by
\begin{equation}\label{eq:bridge_function}
h(\theta)\propto \frac{2}{\eta_1(\theta)+r\eta_2(\theta)},
\end{equation}
see \cite[Theorem 1]{meng1996simulating}. Of course, the fact that $r$ appears in the bridge function expression is problematic. A practical solution is to define a sequence $(r_l)_{l\geq0}$ recursively as
 $$
r_{l} = \sum_{j = 1}^N\frac{\eta_1\left(\theta^{(2)}_j\right)}{\eta_1(\theta^{(2)}_j)+r_{l-1}\eta_2(\theta^{(2)}_j)}\bigg/\sum_{j = 1}^N\frac{\eta_2\left(\theta^{(1)}_j\right)}{\eta_1(\theta^{(1)}_j)+r_{l-1}\eta_2(\theta^{(1)}_j)}\text{, }l\geq1,
$$ 
with some initial value $r_0$. The algorithm stops as soon as the difference between two consecutive $r$ is smaller than some threshold. For our purpose, we set $p_1(\theta) = p\left(\theta  \mid y, M_{k} \right)$ for $k\in\{1,\ldots, K\}$ and therefore $Z_1 = p\left(y \mid M_{k}\right)$. A sample $\theta^{(1)}_1,\ldots,\theta_N^{(1)}\sim p\left(\theta  \mid y, M_{k} \right)$ is readily available from HMC sampling. A common choice for the second distribution $p_2(\theta)$ is the multivariate normal distribution with mean and covariance matrix estimated from the posterior draws, see \cite{overstall2010default} and \cite{gronau2017tutorial}. The bridge sampling algorithm has been implemented in the R package \textbf{bridgesampling}, see \cite{bridgesampling}. \textcolor{black}{Among several importance sampling estimators, \cite{meng1996simulating} showed that the bridge sampler minimizes the mean-squared error and is more robust to the tail behavior of the proposal distribution relative to the posterior distribution \citep{gronau2017tutorial}.} Once the MLs are obtained for each model, weights are given by the posterior model probabilities in Equation \eqref{weightsbma}.	
}
\subsection{Bayesian model averaging by stacking and Pseudo-BMA}\label{sec::BMA2}

Bayesian model averaging is flawed in a setting where the ``true" data-generating process is not part of the model candidates, see \cite{yao2018using}. Indeed, in this setting, BMA asymptotically selects the model in the list which is closest to the real model in the sense of Kullback - Leibler (KL) divergence. More importantly, as we can see from Equation \eqref{ML}, that the marginal likelihood is strongly sensitive to the specific prior choice $p\left(\theta_{k} \mid M_{k}\right)$ in each model, see \cite{fernandez2001benchmark}. 

As an alternative approach, different authors considered model selection and averaging based on prediction performance on hold-out data. For instance, \cite{geisser1979predictive} proposed to replace marginal likelihoods $p\left(y \mid M_{k}\right)$ with a product of Bayesian leave-one-out cross-validation (LOO-CV) predictive densities $\prod_{i=1}^{n} p\left(y_{i} \mid y_{-i}, M_{k}\right)$ where $y_{-i}$ is the data without the $i$-th-point. More recently, \cite{yao2018using} proposed Bayesian model averaging approaches based on LOO-CV. Roughly speaking, weights are chosen such that the averaged model has the best prediction performance according to a logarithm scoring rule. 

In this section, we consider two Bayesian model averaging techniques from \cite{yao2018using}, namely \textit{stacking} and \textit{Pseudo-BMA}, but adapted to the problem of forecasting mortality. As pointed out by \cite{burkner}, LOO-CV is problematic if the goal is to estimate the predictive performance for future time points. Leaving out only one observation at a time will allow information from the future to influence predictions of the past (i.e., data from times $t + 1, t + 2, \dots,$ would inform predictions for time $t$). Instead, it is more appropriate to use leave-future-out validation. In our context of mortality forecasting, instead of leaving one point out, we leave the last $M$ years of data out and evaluate the prediction performance over these $M$ years. 

More precisely, assume that the data for $T$ years is split into a training set and a validation set as follows:
\begin{itemize}
\item $y_{1:N}=(d_{x,t},e_{x,t})$ for all $x$'s and $t=t_1,\dots,t_N$ are the death and exposure counts of the first $N$ years, used to fit the model. 
\item $y_{N+1:N+M}=(d_{x,t},e_{x,t})$ for all $x$'s and $t=t_{N+1},\dots,t_{N+M}$ are the death and exposure counts associated to the remaining $M$ years, used to validate the model. 
\end{itemize}
After fitting the NB model to $y_{1:N}$, we can obtain an empirical distribution of future $\mu_{x,t}$ for $t=t_{N+1},\dots,t_{N+M}$ based on MCMC samples. Combined with the exposures of the validation set, we can then obtain an empirical distribution of future deaths for each model $M_k$: 
\[
D_{x,t}\sim\PoissonDist(\mu^{k}_{x,t}\cdot e_{x,t})
\] 
where $\mu^{k}_{x,t}$ are the forecasted mortality rates under model $M_k$ and $e_{x,t}$, for $t=t_{N+1},\dots,t_{N+M}$, are the exposures of the validation set. A good averaging approach should aggregate the models such that the resulting model maximizes the likelihood of the observed number of deaths on the validation set. This is the key idea of the stacking of predictive distributions. 

\subsubsection{Stacking of predictive distributions}

The first quantity to determine is the posterior predictive density of future deaths given the training data, i.e. $p(d_{x,j}|y_{1:N})$ for all validation years $j=t_{N+1},\dots,t_{N+M}$. These quantities can be computed with the help of the posterior distribution $p\left(\theta \mid y_{1:N}\right)$ of the parameters $\theta$ conditionally to the training dataset for each model $M_k$. Formally, we have
\begin{equation}\label{postdensity}
p\left(d_{x,j} \mid y_{1: N},M_k\right)=\int p\left(d_{x,j} \mid y_{1: N}, \theta,M_k\right) p\left(\theta \mid y_{1: N},M_k\right) \mathrm{d} \theta.
\end{equation}
The density \eqref{postdensity} is analytically intractable but can be approximated based on MCMC samples. Having obtained $S$ draws $\left(\theta^{(1)}, \ldots, \theta^{(S)}\right)$ from the posterior distribution $p\left(\theta \mid y_{1:N},M_k\right)$, we simply approximate $p\left(d_{x,j} \mid y_{1: N},M_k\right) $ by 
\begin{equation*}
p\left(d_{x,j} \mid y_{1: N},M_k\right)\approx \frac{1}{S} \sum_{s=1}^{S} p\left(d_{x,j} \mid y_{1: N}, \theta^{(s)},M_k\right).
\end{equation*}
The goal of stacking a set of $K$ predictive distributions built from the models $\mathcal{M}=\left(M_{1}, \ldots, M_{K}\right)$ is to find the distribution in the convex hull $\mathcal{C}=\left\{\sum_{k=1}^{K} w_{k} \times p\left(\cdot \mid M_{k}\right): \sum_{k} w_{k}=1, w_{k} \geq 0\right\}$ that is optimal according to some given criterion. In this paper, we follow the approach of \cite{yao2018using} and use a logarithm scoring rule to define the optimality criterion. The weights $w_k,\, k= 1,\ldots, K,$ associated to each mortality model $M_k\in \mathcal{M}$ follows from solving the optimization problem
\begin{equation*}
\max _{w \in \mathcal{S}_{1}^{K}} \sum_{x=x_1}^{x_n}  \sum_{j=t_{N+1}}^{t_{N+M}} \log \sum_{k=1}^{K} w_{k} p\left(d_{x,j} \mid y_{1: N},M_k\right), 
\end{equation*}
where
$$
\mathcal{S}_{1}^{K}=\left\{w \in[0,1]^{K}: \sum_{k=1}^{K} w_{k}=1\right\}.
$$
The combined predictive distribution is then given by 
\begin{equation*}
p\left(d_{x,j} \mid y_{1: N}\right)= \sum_{k=1}^{K} w_{k}  p\left(d_{x,j} \mid y_{1: N},M_k\right).
\end{equation*}
By construction, this averaged distribution maximizes the log likelihood of the observed number of deaths in the validation set among all distributions in the convex hull $\mathcal{C}$.

\subsubsection{Pseudo-BMA}
As an alternative approach, we consider an AIC-type weighting scheme using leave-future-out validation. To compare the different models, we use the expected log predictive density for each model $M_k$ ($\operatorname{elpd}^{k}$) as a measure of predictive accuracy, see \cite{vehtari2017practical}. The $\operatorname{elpd}^{k}$ is defined as follows:
\begin{equation}
\operatorname{elpd}^{k}= \sum_{x=x_1}^{x_n}  \sum_{j=t_{N+1}}^{t_{N+M}} \log p\left(d_{x,j} \mid y_{1: N},M_k\right).
\end{equation}
Hence, $\operatorname{elpd}^{k}$ is the sum of the point-wise posterior predictive densities over all held-out data points, namely observed deaths for all ages $x$ and all validation years $j=t_{N+1},\dots,t_{N+M}$. We can interpret $\operatorname{elpd}^{k}$ as an aggregate measure of how well the model $M_k$ predicts the observed deaths in the validation set. The Pseudo-BMA weight for model $M_k$ is given by 
\begin{equation*}
w_{k}=\frac{\exp \left(\operatorname{elpd}^{k}\right)}{\sum_{k=1}^{K} \exp \left(\operatorname{elpd}^{k}\right)}.
\end{equation*}

\section{Simulation study}\label{sec:sim} 
A simulation experiment is carried out in order to better understand the behaviour of the selection methods described in \autoref{sec:BMMA}. We take the Belgian mortality data for calendar years from $1959$ to $2019$ and people aged $50$ to $90$. A mortality model is fitted to these data and the draws from the posterior distribution are then used to generate $80$ synthetic mortality data sets. The dimensions of the synthetic data corresponds exactly to the original mortality data. The various mortality models are fitted to the synthetic datasets, the last ten calendar years of which have been set aside for the evaluation of the out-of-sample forecast error. We want to measure the ability of the selection method to choose the most suitable model. \textcolor{black}{ We do this by inspecting the value of the weights returned by each method. We assess the predictive power of the model averaging strategies by examining how well the predicted mortality rates $\mu_{x,t}$ overlap with the mortality rates of the test set 
$$
y = (d_{x,t}, e_{x,t}),\text{ for }x = 50,\ldots, 90\text{ and }t = 2010,\ldots, 2019.
$$
Because we use Bayesian inference, we have a probability distribution $F_{x,t}$ around each mortality rate $\mu_{x,t}$. The accuracy of this family of forecast distributions $F = (F_{x,t})$ is measured via two scoring rules. The logarithmic score is defined as 
\begin{equation}\label{eq:logS}
\operatorname{LogS}(F, y)=-\frac{1}{40}\sum_{x=50}^{90}\frac{1}{10}\sum_{t=2010}^{2019}\log\left[f_{x,t}\left(\frac{d_{x,t}}{e_{x,t}}\right)\right],
\end{equation}
where $f_{x,t}$ is the PDF of $F_{x,t}$. The continuous ranked probability score (CRPS) is given by
\begin{equation}\label{eq:CPRS}
\operatorname{CRPS}(F, y)=\frac{1}{40}\sum_{x=50}^{90}\frac{1}{10}\sum_{t=2010}^{2019}\int_{\mathbb{R}}(F_{x,t}(\mu)-\mathds{1}\{d_{x,t}/e_{x,t} \leq \mu\})^{2} \mathrm{~d} \mu.
\end{equation}
The evaluation of the criteria \eqref{eq:logS} and \eqref{eq:CPRS} requires to replace the CDFs $F_{x,t}$ and the PDFs $f_{x,t}$ by their empirical counterparts recovered from our HMC samples. The forecast distributions associated to the averaging methods correspond to a mixture of the forecast distributions associated to the single mortality models. The use of such scoring rules to compare probabilistic population forecasts is discussed in the work of \citep{Keilman2020}. The concrete evaluation of the score is done using the R package \textbf{scoringRules} \citep{jordan}. Finally, the pointwise accuracy of the model averaging strategies is measured by the mean absolute error (MAE) of the posterior means $\widehat{\mu}_{x,t}$, based on the test dataset, averaged over all ages as}
\begin{equation}\label{eq:mae_sim_study}
\text{MAE}=\frac{1}{40}\sum_{x=50}^{90}\frac{1}{10}\sum_{t=2010}^{2019} \abs{d_{x,t}-e_{x,t}\widehat{\mu}_{x,t}}. 
\end{equation}
The depth of the data history ranges from $20$ up to $50$ calendar years. For the pseudo-BMA and stacking approach, we have considered validation sets containing $1, 5$ and $10$ calendar years. The split of the data between training, validation and test sets is summarized in \autoref{tab:simulation_study_time_line}.
\begin{table}[h]
\centering
\begin{tabular}{llccc}
\toprule
	&& Fitting    & Validation     & Prediction                   \\ 
	\midrule
	BMA&& 1989-2009 & & 2010-2019\\
	&& 1979-2009 & &-\\
	&& 1969-2009 & &-\\
	&& 1959-2009 & &-\\ 	
	\midrule	
	pseudo-BMA / Stacking&& 1989-2008 &2009 &2010-2019 \\
	&& 1989-2004 &2005-2009 & -  \\
	&& 1989-1999 &2000-2009 & -  \\
	\cmidrule{3-5}
	&& 1979-2008 &2009 &2010-2019 \\
	&& 1979-2004 &2005-2009 & -  \\
	&& 1979-1999 &2000-2009 & -  \\
	\cmidrule{3-5}
	&& 1969-2008 &2009 &2010-2019 \\
	&& 1969-2004 &2005-2009 & -  \\
	&& 1969-1999 &2000-2009 & -  \\
	\cmidrule{3-5}
	&& 1959-2008 &2009 &2010-2019 \\
	&& 1959-2004 &2005-2009 & -  \\
	&& 1959-1999 &2000-2009 & -  \\

		\bottomrule
	\end{tabular}
\caption{Simulation experiments and time-line assumptions}\label{tab:simulation_study_time_line}
\end{table}
We have noticed that a validation set containing only one calendar year is insufficient, hence the results are not reported for brevity. The difference between having $5$ or $10$ years in the validation set is so small that we only report the results associated with a validation set containing $10$ years of data. Note that it is consistent with the size of the test dataset. We consider two cases:
\begin{itemize}
	\item  In the first one, the data is generated by an Age-Period-Cohort model. The true model is among the competing models and the results are given in \autoref{ssec:apc_synthetic}.
	\item In the second case, the numbers of death result from taking the average of death counts drawn from a Cairns-Blake-Dowd model and from a Renshaw-Haberman model. The true model is not among the competing models, which brings us closer to a real situation. The results are discussed in \autoref{ssec:mix_cbd_rh_synthetic}.      
\end{itemize}
\subsection{Data generated by an Age-Period-Cohort model}\label{ssec:apc_synthetic}
\begin{figure}[h!]
	\begin{center}
		\subfloat[$\alpha_x,\text{ }x = 50\ldots, 90$]{
			\includegraphics[width=0.4\textwidth]{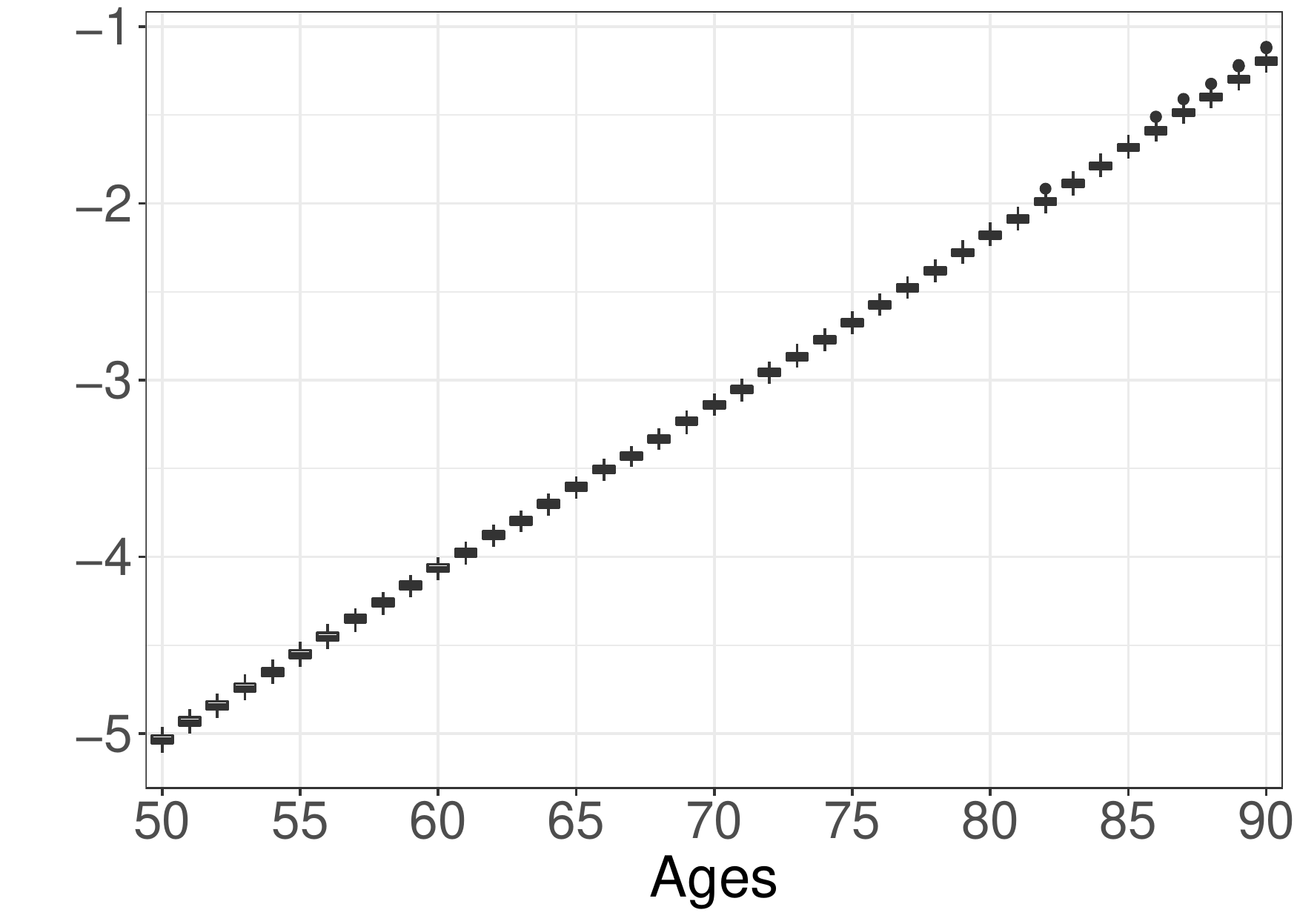}
			\label{sub:apc_a_plot}
		}
		\hskip1em
		\subfloat[$\kappa_t,\text{ }t = 1960\ldots, 2019$]{
			\includegraphics[width=0.4\textwidth]{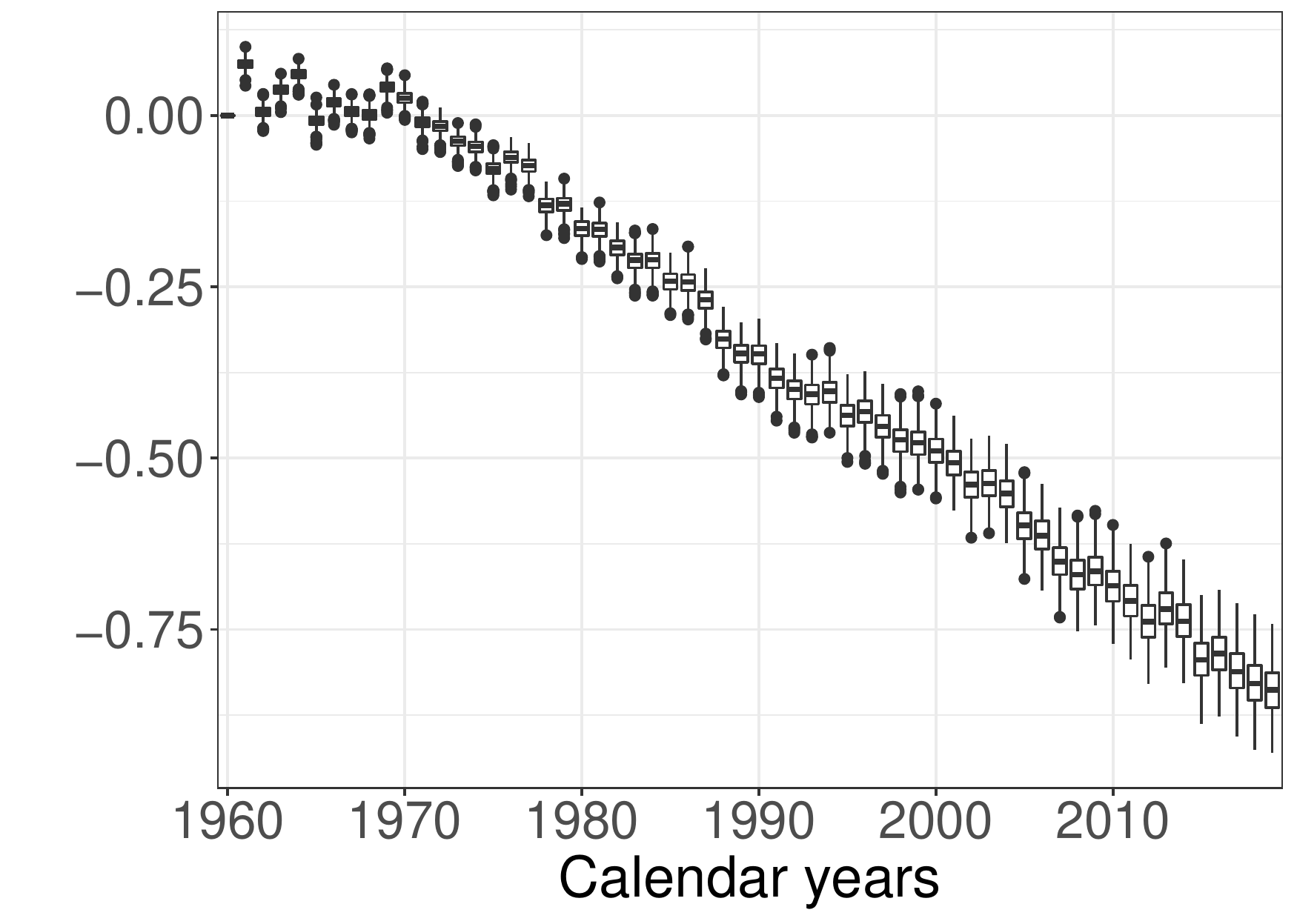}
			\label{sub:apc_k_plot}
		}
		\hskip1em
		\subfloat[$\gamma_{t-x}$, $x = 50\ldots, 90;t = 1960\ldots, 2019$]{
			\includegraphics[width=0.4\textwidth]{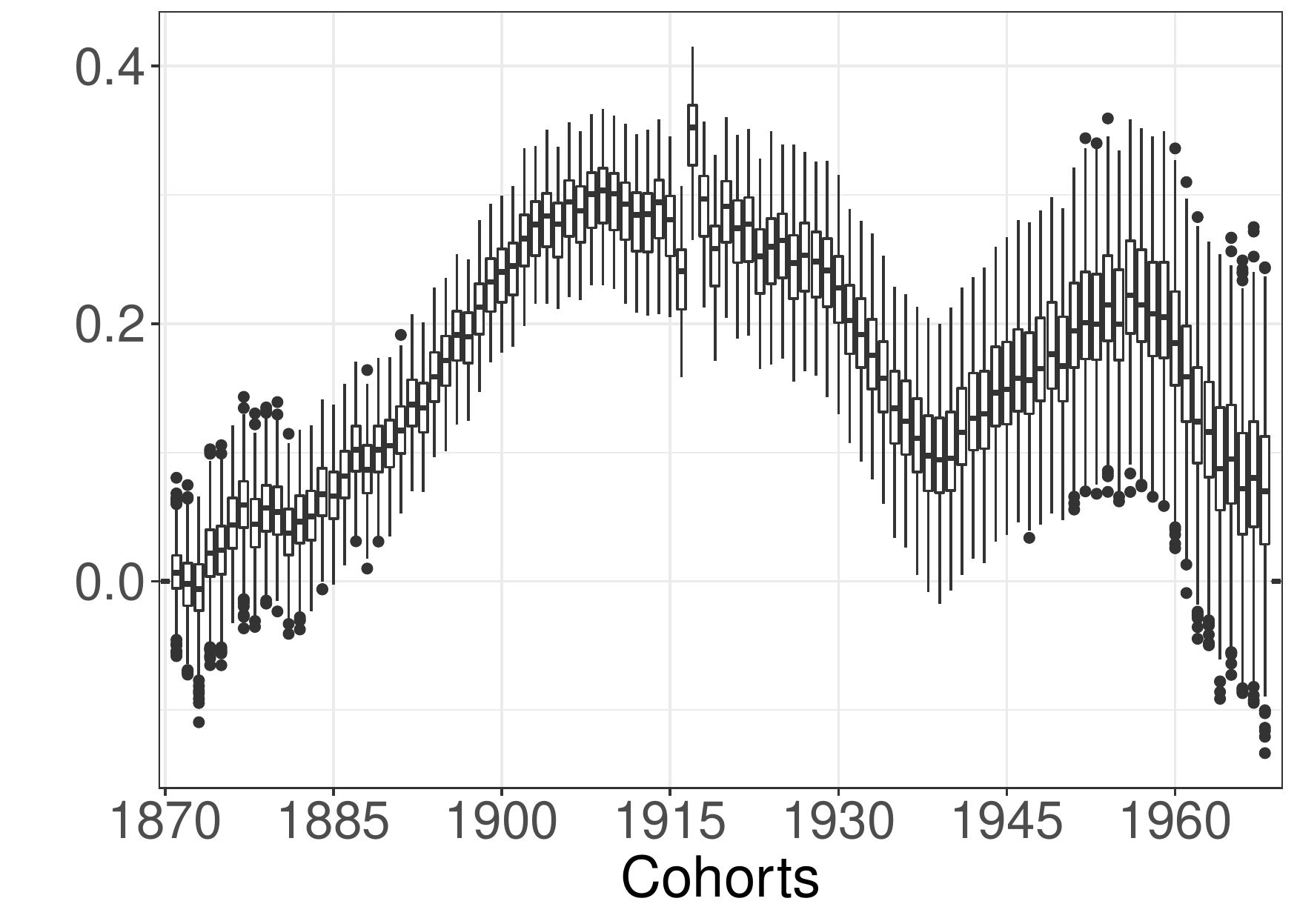}
			\label{sub:apc_g_plot}
		}
		\hskip1em
		\subfloat[$\phi$]{
			\includegraphics[width=0.4\textwidth]{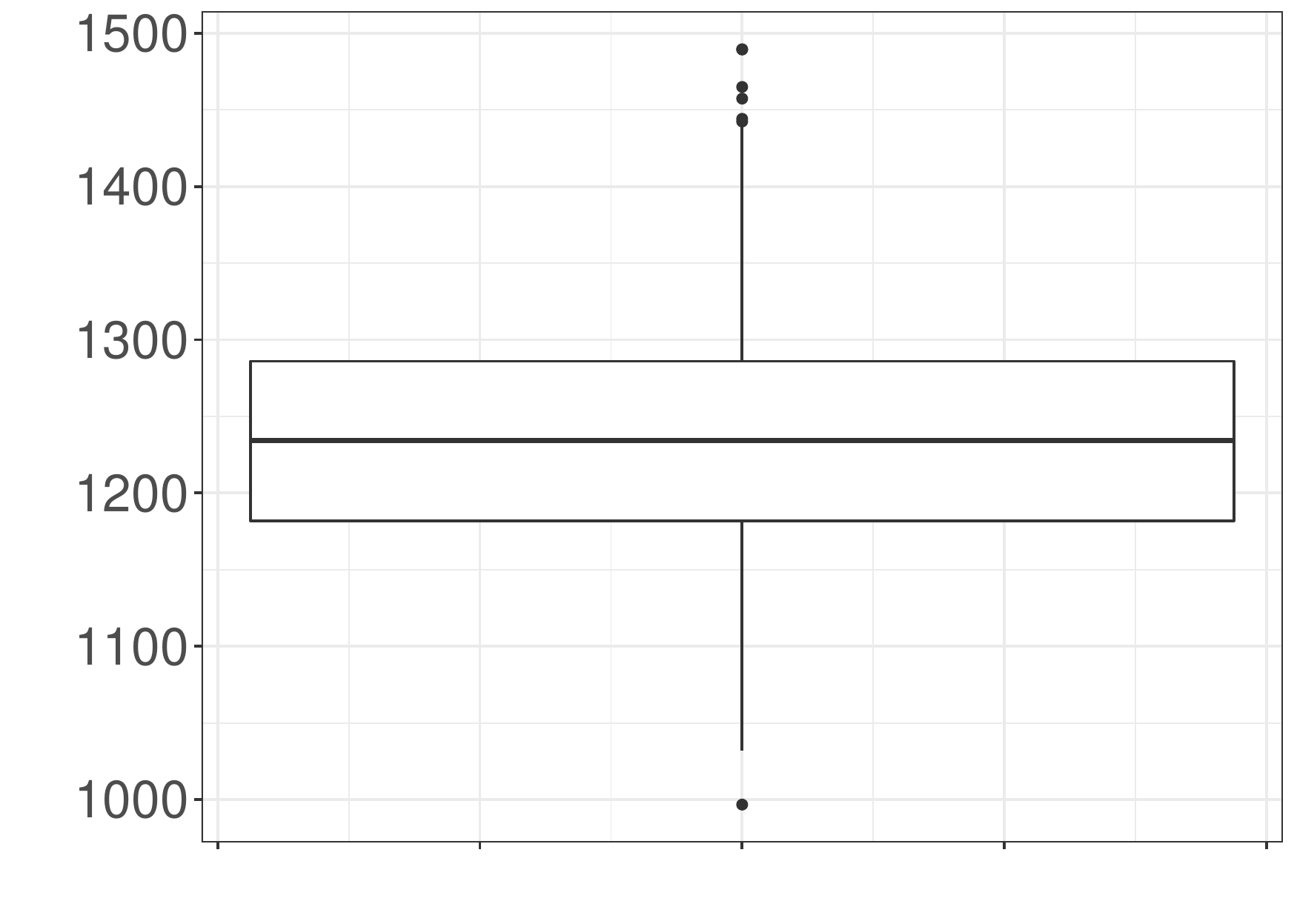}
			\label{sub:apc_phi_plot}
		}
		\caption{Posterior distribution of the parameters of the APC model that generated the synthetic data.}
		\label{fig:apc_model_parms}

	\end{center}
\end{figure}

\begin{figure}[h!]
	\begin{center}
		\includegraphics[scale=0.45]{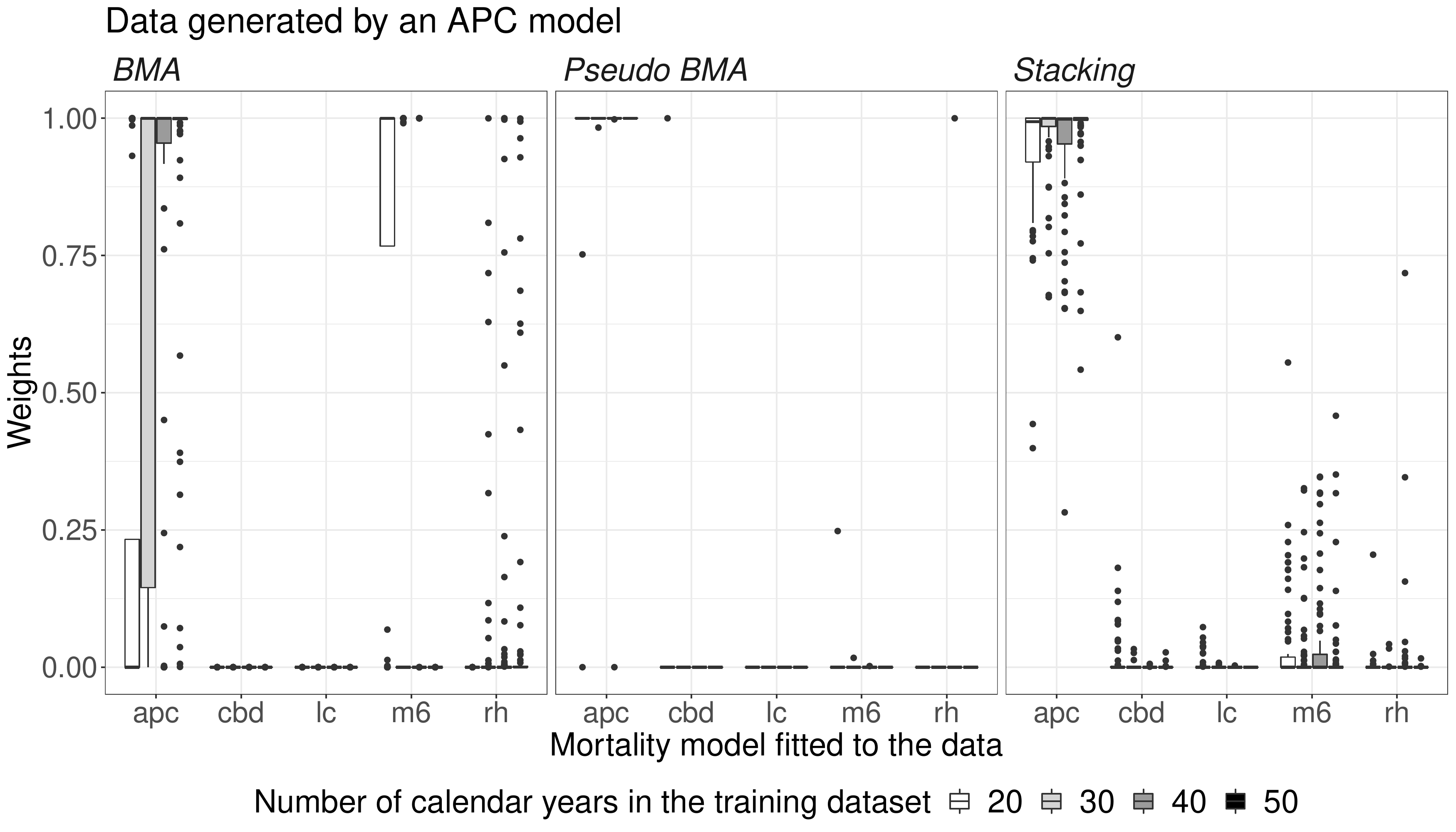}
		\caption{Weights assigned to each mortality model, depending on the method and the number of calendar years in the training set for $80$ synthetic data sets generated by an Age-Period-Cohort model.}
		\label{fig:weights_plot_apc}
	\end{center}
\end{figure}

\begin{figure}[h!]
	\begin{center}
		\includegraphics[scale=0.45]{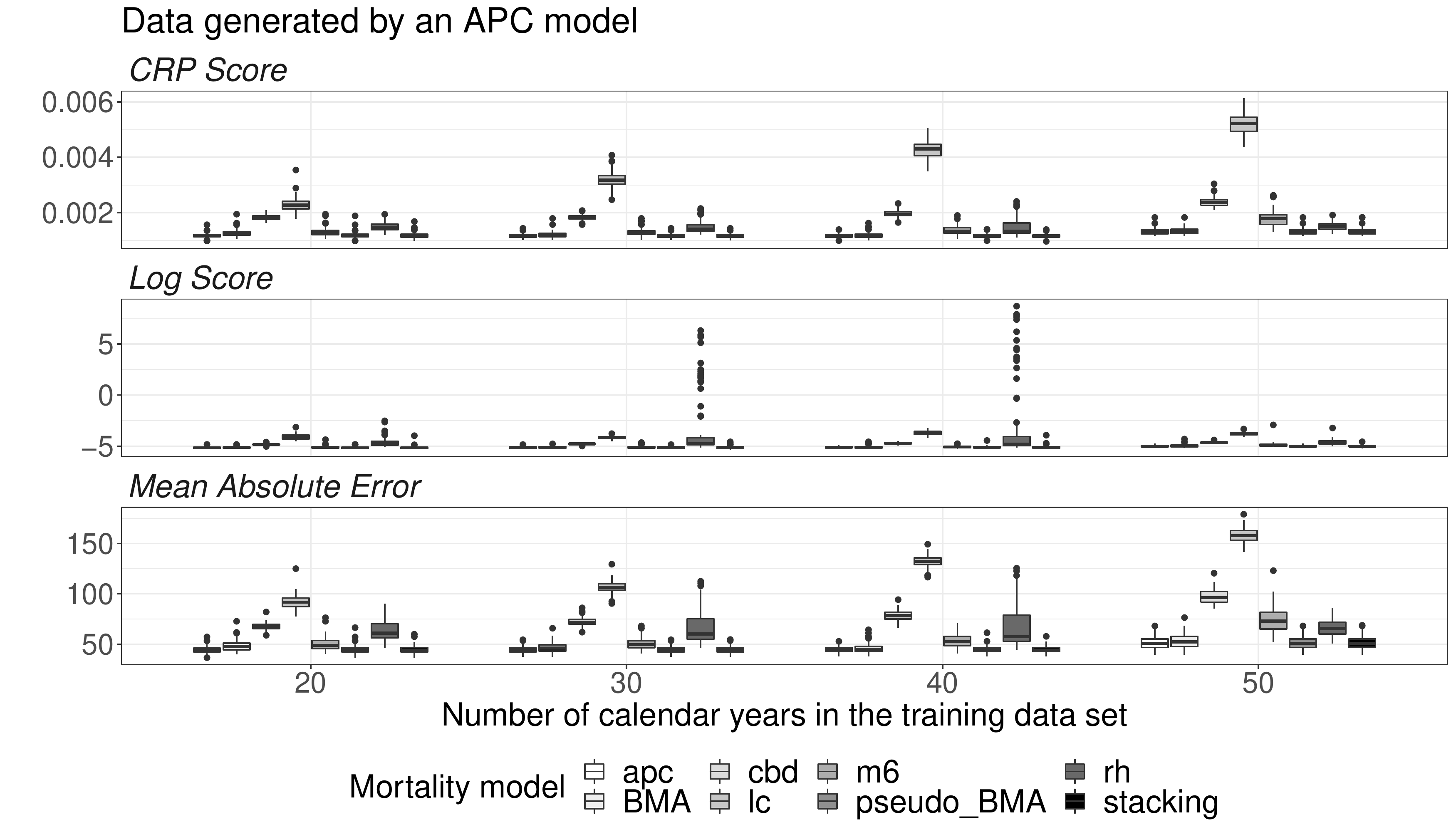}
		\caption{Logarithmic score, CRP score and mean absolute errors calculated over $80$ simulated datasets from an APC model depending on the number of calendar years in the training dataset.}
		\label{fig:mae_plot_apc}
	\end{center}
\end{figure}

The APC model is fitted to the Belgian mortality data for calendar years from $1959$ to $2019$ and people aged $50$ to $90$, the posterior distribution of the parameters is provided on Figure \autoref{fig:apc_model_parms}.
Based on the posterior draws, $80$ synthetic mortality datasets are generated to which are fitted the mortality models including LC, CBD, APC, RH and M6. The synthetic data provided to the mortality models only contain the calendar years from $1959$ to $2009$, the remaining ten years are kept as a test set to assess the predictive power through the scoring rules and the mean absolute error defined in \eqref{eq:logS}, \eqref{eq:CPRS}, and \eqref{eq:mae_sim_study}. The validation set for the pseudo-BMA and stacking methods contains $10$ years of data. \autoref{fig:weights_plot_apc} shows the distribution of the weights assigned to each mortality model depending on the averaging method and the number of calendar years in the training dataset.
We note that all the methods discard the CBD and LC models as they do not account for the cohort effect. The pseudo-BMA and stacking methods clearly favor the APC model. The standard BMA approach favors the M6 model when $20$ years are included in the training data set before clearly siding for the APC model. \autoref{fig:mae_plot_apc} displays the logarithmic score, the CRP score, and the mean absolute error of the prediction resulting from the mortality models and their combination via the different methods of model aggregation as a function of the number of calendar years in the training dataset.   
As expected, the best prediction is provided by the APC model while the predictions made by the CBD and LC models are quite flawed. Since stacking and pseudo-BMA tend to always choose the APC model, their use leads to a slight improvement in predictions over the BMA approach. We note that the prediction error is slightly higher when 50 calendar years are included in the training data set. This might seem counter-intuitive as one would expect that the more data there is, the better the prediction. This is not generally true when studying mortality. \textcolor{black}{Years too far from the projection horizon may degrade the forecast, especially if the period effect $\kappa_{t}$ presents structural changes \citep{van2016impact}. In particular, \autoref{fig:apc_model_parms} shows that $\kappa_{t}$ is rather constant between 1960 and 1970 and then decreases after 1970, representing the improvement in longevity from 1970. This, together with \autoref{fig:mae_plot_apc}, suggests that including the data from 1960 to 1970 deteriorates the mortality predictions.}  If $20$ calendar years seem sufficient to make reasonable predictions, taking $30$ or $40$ calendar years widens the gap between the prediction errors resulting from models that encapsulate a cohort effect and those that do not. The case studied in this section corresponds to a situation where the model is well specified because the APC model belongs to the competing models. The next section will allow us to see whether these results are also valid in a misspecified case.
\subsection{Data generated by a mixture between a CBD and RH model}\label{ssec:mix_cbd_rh_synthetic}
\begin{figure}[h!]
	\begin{center}
		\includegraphics[scale=0.45]{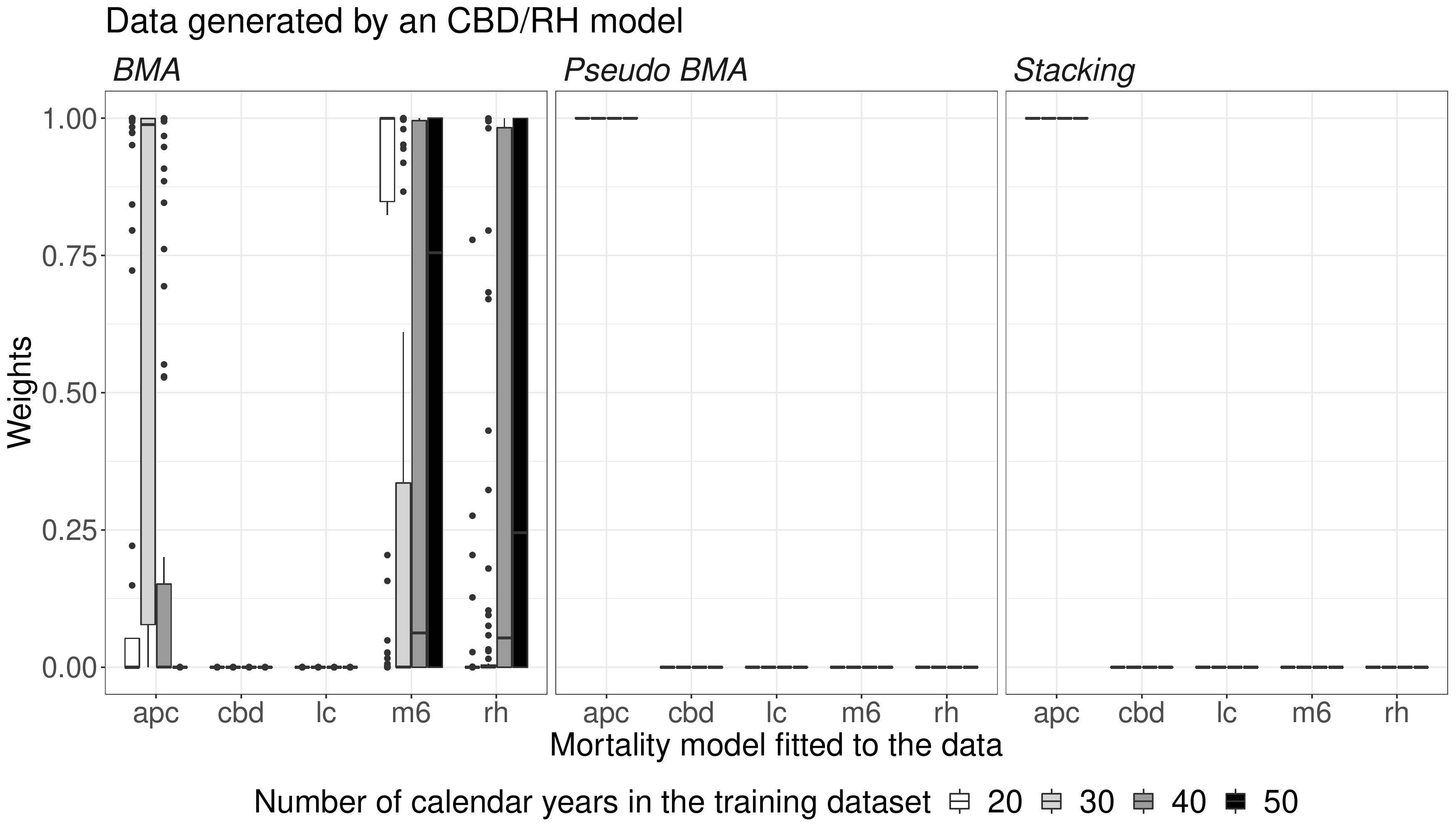}
		\caption{Weights assigned to each mortality model, depending on the method and the number of calendar years in the training set for $80$ synthetic datasets generated by a mixture of a CBD model and a RH model.}
		\label{fig:weights_plot_mix_cbd_rh}
	\end{center}
\end{figure}

\begin{figure}[h!]
	\begin{center}
		\includegraphics[scale=0.45]{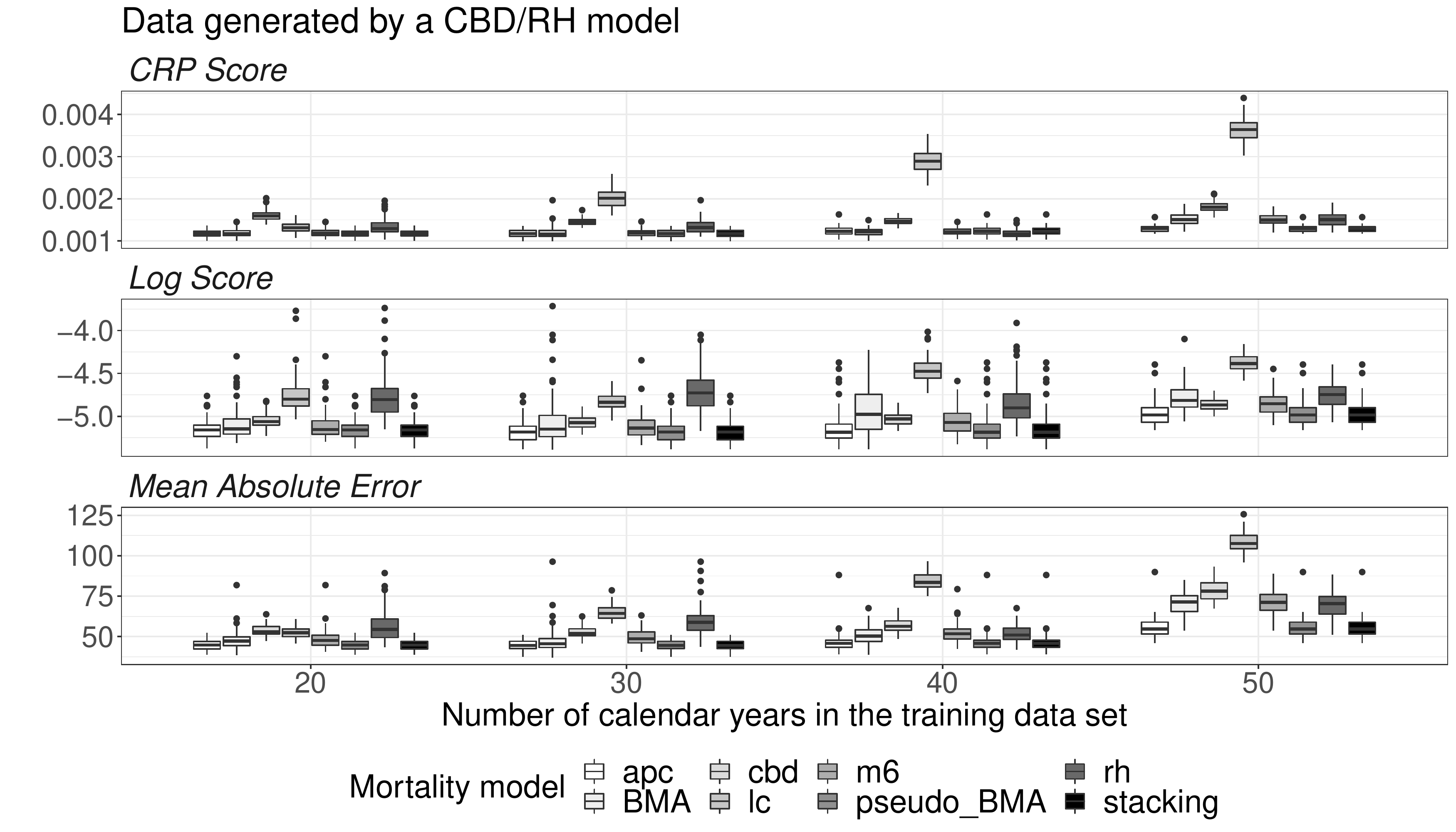}
		\caption{Logarithmic score, CRP score and mean absolute errors calculated over 80 data sets simulated from a mixture of a CBD model and a RH model depending on the number of calendar years in the training data set.}
		\label{fig:mae_plot_mix_cbd_rh}
	\end{center}
\end{figure}

The same Belgian mortality data set is used to fit the CBD and RH models. Each model is used to generate $80$ synthetic mortality data sets. The synthetic data sets are combined in pairs by taking the average number of deaths. We then fit the mortality models to these hybrid mortality data (without the last ten years that will be used to measure the out-of-sample error) and apply the different models averaging strategies. \autoref{fig:weights_plot_mix_cbd_rh} shows the distribution of the weights assigned to each mortality model depending on the averaging method and the number of calendar years in the training dataset.
The BMA approach favors the M6 model but also chooses from time to time the RH and APC models. The stacking and pseudo-BMA techniques clearly side for the APC model. Let us see what it means in terms of the prediction errors. \autoref{fig:mae_plot_mix_cbd_rh} displays the logarithmic score, the CRP score, and the mean absolute error of the prediction resulting from the mortality models and their combination via the different methods of model aggregation depending on the number of calendar years in the training datasets.   
The APC model returns the smallest prediction error and the same goes for stacking and pseudo-BMA approaches which tend to give the APC model a lot of credibility. Again, taking $50$ calendar years is detrimental to the accuracy of the forecast. This study demonstrates the good behavior of the Bayesian model averaging methods in a controlled environment (the data generation process being specified by us). Due to the good performances of the APC model, and the fact that the selection methods allocate a weight close to $1$ to it prevents the model averaging methods from making better prediction than that the APC model.
The following section is devoted to the application to actual mortality data sets.

\section{Application to real mortality data}\label{sec:real_data}

In this section we apply the three model averaging approaches discussed in \autoref{sec:BMMA} to mortality data from France, UK, USA and Japan. The data chosen for illustrative purposes are the male death data and the corresponding exposures of these four countries, for ages $50-90$ and the last 40 years of data available (1979-2018) extracted from the Human Mortality Database (HMD)\footnote{See \url{www.mortality.org.}}. To assess the prediction performance, we split the data into two parts: the first $30$ years are used for the weights selection (1979-2008) and the last $10$ years (2009-2018) are used to compare the weighted forecasts. For the calculation of the stacking and pseudo-BMA weights, the data is then divided into two parts: the first $20$ years are used as a training set while the remaining $10$ years are used for validation. The size of the leave-future-out validation set is consistent with the findings of \autoref{sec:sim}. The data partitions associated to each model averaging method are given in \autoref{tab:split}.
\begin{table}[h]
	\centering
	\caption{Fitting, validation and prediction periods for the three model averaging approaches.}\label{tab:split}	
	\begin{tabular}{lccc}
\toprule
		& 1979-1998    & 1999-2008     & 2009-2018                   \\ 
		\midrule
		\multicolumn{1}{l}{BMA}        & \multicolumn{2}{c}{Fitting} & \multirow{3}{*}{Prediction} \\ 		\cmidrule{1-3}
		\multicolumn{1}{l}{Stacking}   & Fitting      & Validation    &                             \\ 
		\cmidrule{1-3}
		\multicolumn{1}{l}{Pseudo-BMA} & Fitting      & Validation    &                             \\ 
		\bottomrule
	\end{tabular}
\end{table}

\subsection{Model Weights}

\autoref{weightcountries} provides the weights obtained via standard BMA (marginal likelihood), stacking and pseudo-BMA for France, UK, USA and Japan. 
\begin{table}[h!]
	\centering
	\caption{Model Weights for France, UK, USA and Japan via BMA, stacking and pseudo-BMA.} \label{weightcountries}
	\begin{tabular}{lccccccc}
		\toprule
		\multicolumn{1}{c}{\multirow{2}{*}{}} & \multicolumn{3}{c}{France}                                         & \multicolumn{1}{l}{} & \multicolumn{3}{c}{UK}                                             \\
		\cmidrule{2-4}\cmidrule{6-8}
		\multicolumn{1}{c}{}                       & BMA                  & Stacking             & Pseudo-BMA           &                      & BMA                  & Stacking             & Pseudo-BMA           \\
		\midrule
		LC                                         & 0                    & 0.093                & 0                    &                      & 0                    & 0                    & 0                    \\
		RH                                         & 1                    & 0.750                    & 1                    &                      & 0                    & 0.298                    & 0                    \\
		APC                                        & 0                    & 0.157                & 0                    &                      & 0                    & 0                    & 0                    \\
		CBD                                        & 0                    & 0                    & 0                    &                      & 0                    & 0                & 0                    \\
		M6                                         & 0                    & 0                    & 0                    &                      & 1                    & 0.702                & 1                    \\
		\midrule
		\multirow{2}{*}{}                          & \multicolumn{3}{c}{USA}                                            & \multicolumn{1}{l}{} & \multicolumn{3}{c}{Japan}                                          \\
				\cmidrule{2-4}\cmidrule{6-8}
		& BMA                  & Stacking             & Pseudo-BMA           &                      & BMA                  & Stacking             & Pseudo-BMA           \\
		\midrule
		LC                                         & 0                    & 0                & 0                    &                      & 0                    & 0.367                & 0                    \\
		RH                                         & 1                    & 0.71                    & 0.982                    &                      & 1                    & 0.174                & 0                    \\
		APC                                        & 0                    & 0.29                & 0.018                    &                      & 0                    & 0.458                & 1                    \\
		CBD                                        & 0                    & 0                & 0                    &                      & 0                    & 0                    & 0                    \\
		M6                                         & 0                    & 0                    & 0                    &                      & 0                    & 0                    & 0          \\         
\bottomrule
	\end{tabular}
\end{table}
 The BMA and pseudo-BMA approaches tend to only select one model. This was expected given the size of the dataset (see \cite{yao2018using} and the references therein). On the other hand, the stacking approach selects two models for UK and USA and three models for France and Japan. Overall, we observe a certain agreement between the stacking and pseudo-BMA approaches based on validation while the BMA and pseudo-BMA do not always select the same model. We also note that the standard BMA approach favors either the RH model or the M6 model\footnote{We note that the model selection via BMA is sensitive to the sample period used to fit the models. For a 20-year fitting period (1979-2008), we found that the M6 model was selected for France and UK, and the APC model for USA and Japan. The sensitivity of mortality models to the sample period has been extensively studied and we refer to \cite{cairns2011mortality} among others.}; this is in agreement with the frequentist literature in which the RH model or the CBD with cohort effect have been often identified as the best candidate model when model selection is based on the BIC or AIC criterion, see \cite{cairns2009quantitative} and \cite{haberman2011comparative} among others.
%

\subsection{Prediction performance}
\begin{sidewaysfigure}[tbp]
		\includegraphics[width=\linewidth]{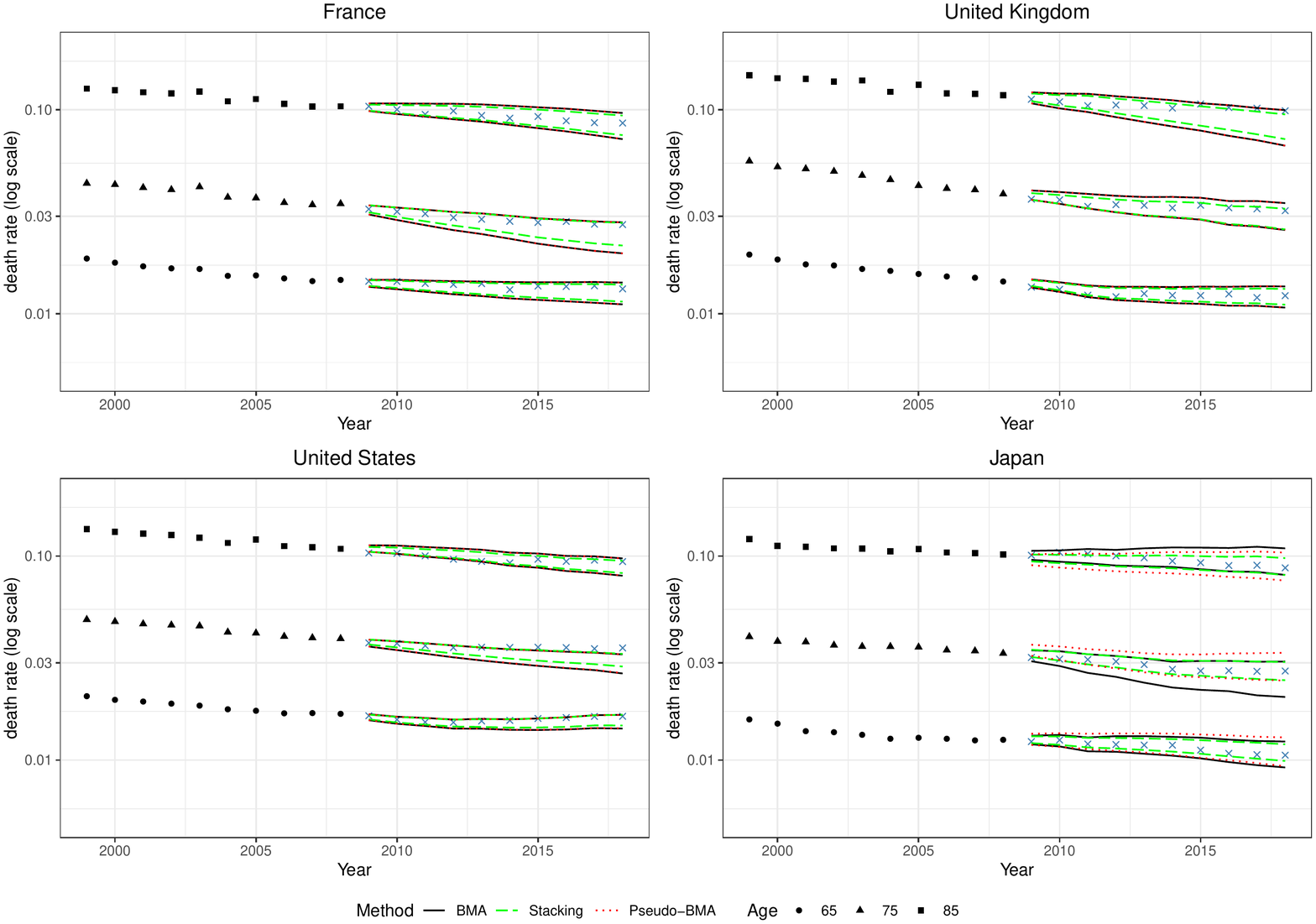} 
		\caption{95\% prediction intervals for the death rates for age $x=65,75,85$ via the three model averaging approaches along with the observed crude deaths rates from France, UK, USA and Japan for the 10-year period 2009-2018.}
		\label{predictionintervals}
\end{sidewaysfigure}

To assess the prediction performance of the three Bayesian model averaging approaches, we first compute the $95\%$ credible intervals of the projected log death rates for age $x=65,75,85$ as a function of time, $10$ years into the future, along with the observed crude death rates as shown in \autoref{predictionintervals}. An ideal credible interval should be sufficiently large to contain the observed death rates of the next $10$ years but not too wide to avoid overconservative credible intervals. We note the following: 
\begin{itemize}
\item For France, the three methods provide reasonable and similar credible intervals at age $85$. However, at age $75$ and age $65$, whatever the approach, the intervals seem to be too narrow as the last death rates tend to fall outside the confidence bands.
\item For the UK, we also observe that the intervals are too narrow at age $85$ while the observed death rates fall right inside the intervals at age $65$ and $75$. 
\item For the USA, the model averaging methods fail to match the observed death rates at age $75$ as they lie outside the confidence interval.
\item  \textcolor{black}{For Japan, we observe that for the ages 65 and 75, the observed death rates are more centered for the stacking approach while standard BMA better projects at age 85 as credible intervals encompass the observed death rates at that age.}
\end{itemize}

\begin{sidewaysfigure}[tbp]
		\includegraphics[width=\linewidth]{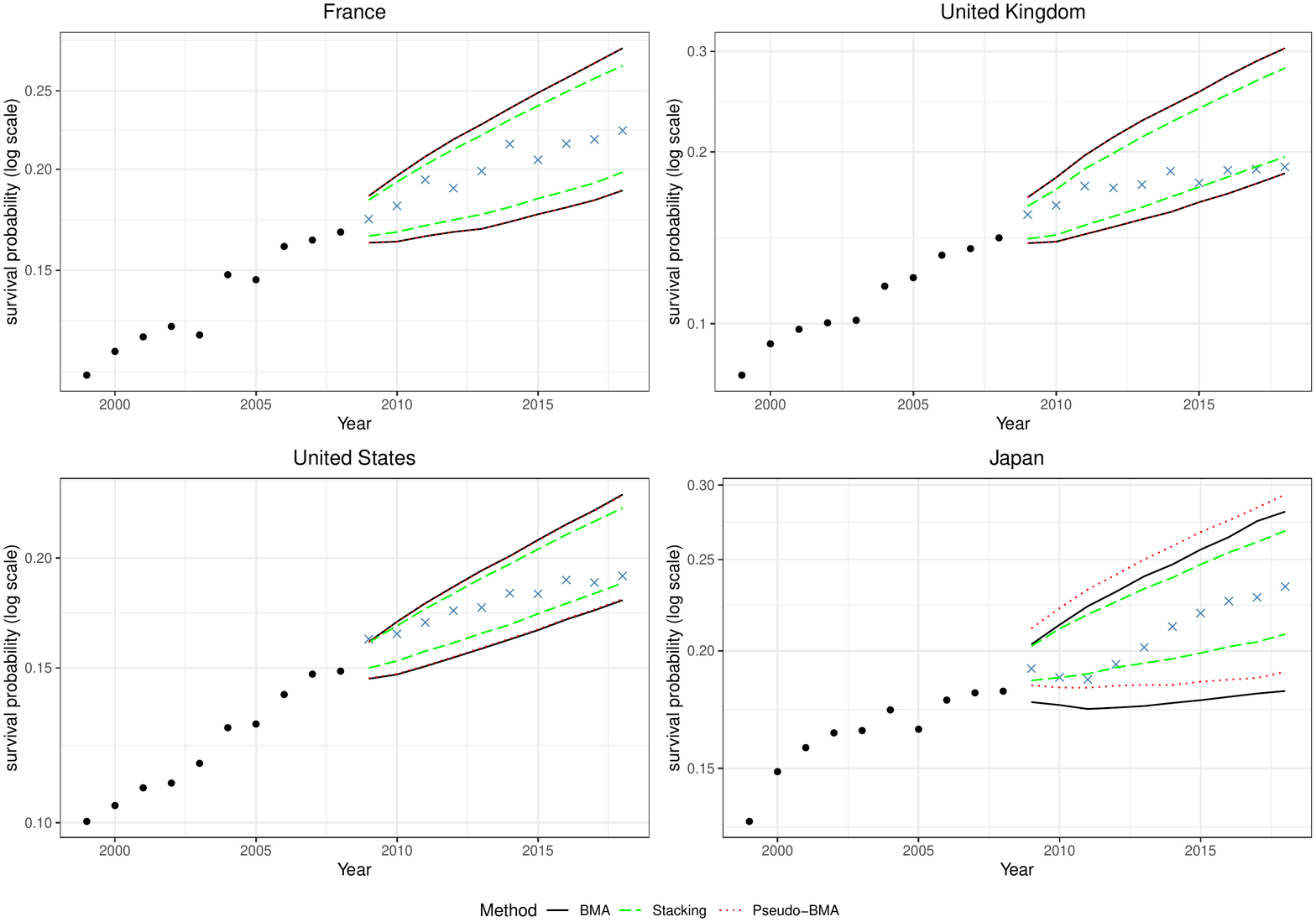} 
		\caption{95\% prediction intervals for period survival probability at age 50 until 90 via the three model averaging approaches along with the observed period survival probabilities from France, UK, USA and Japan for the 10-year period 2009-2018.}
		\label{survprob}
\end{sidewaysfigure}

\begin{sidewaysfigure}[tbp]
		\includegraphics[width=\linewidth]{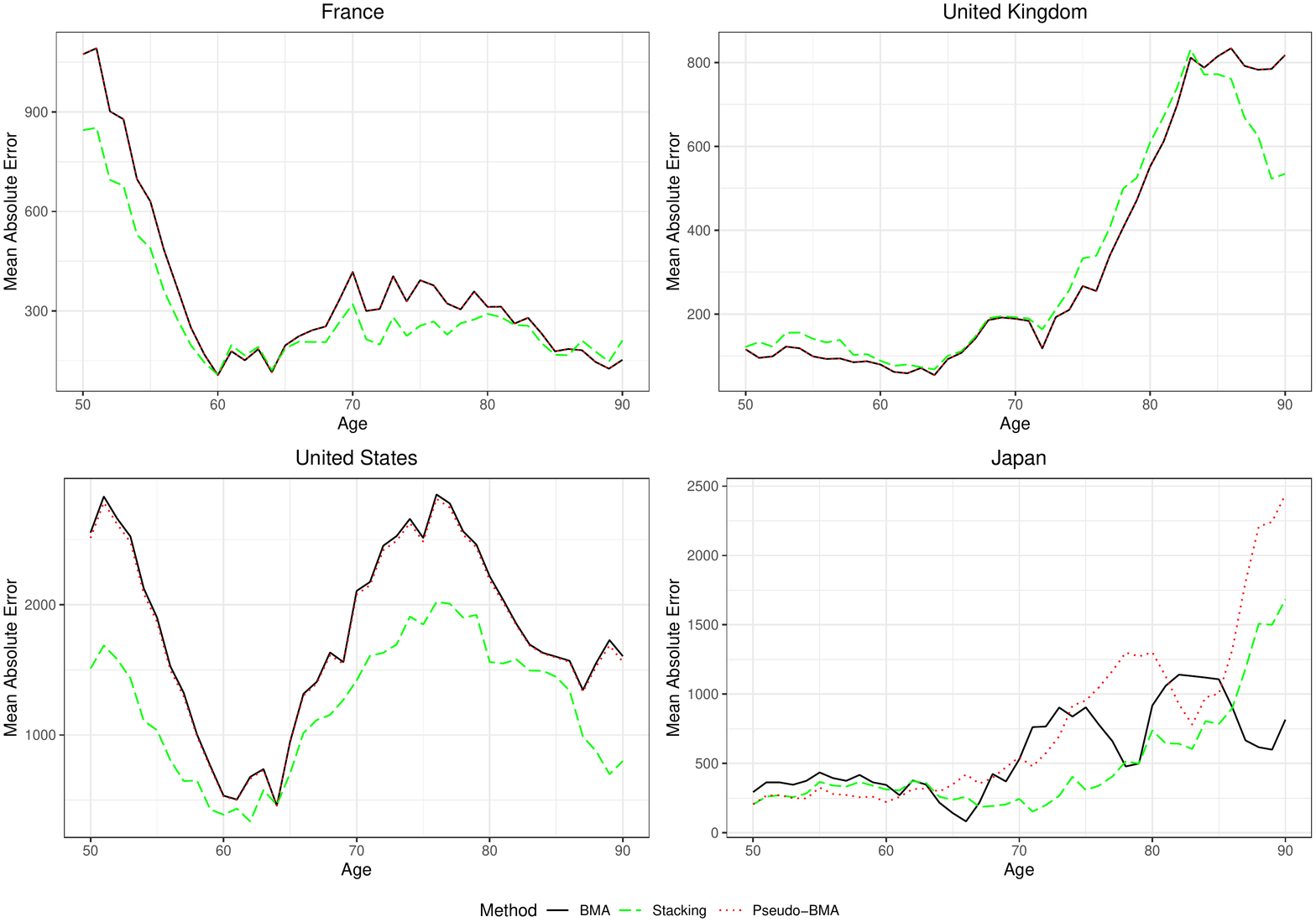} 
		\caption{Mean Absolute Error per age (50-90) averaged across years (2009-2018).}
		\label{mae}
\end{sidewaysfigure}

We now study the performance of the models when estimating mortality indicators that aggregate all ages. A common quantity is the life expectancy at birth but it would require the full age range. Since we focus on the age range $50-90$, we instead compute a $40$-year period survival probability of a person of age $x=50$ for any year $t$: 
\begin{equation}
_{40}p_{50,t}=\prod_{i=0}^{39}p_{50+i,t}=\prod_{i=0}^{39}\exp\left(-\mu_{50+i,t}\right).\label{survivalprobability}
\end{equation}
It corresponds to the probability that a $50$ years old person to live for more than $40$ additional years given the mortality conditions at year $t$. On \autoref{survprob}, we have plotted the $95\%$ credible intervals of the period survival probabilities for the $10$-year period $2009-2018$, along with the observed quantities. For France, the holdout survival probabilities lie within the $95\%$ prediction intervals of the three model averaging approaches. However, for the UK, the stacking approach overestimates the survival probabilities while the BMA and Pseudo-BMA approaches manage to get the observed quantities in their prediction intervals. For Japan, the intervals obtained by stacking seem to be slightly too narrow as the first holdout points lie outside the prediction intervals. For the four countries considered, the BMA and Pseudo-BMA slightly outperforms the stacking approach by providing wider confidence intervals for the survival probability.

To close, we also assess the predictive performance by age for each country through the Mean Absolute Error (MAE) over the years in the test set: 
\begin{equation*}
\text{MAE}_x=\frac{1}{10}\sum_{t=2009}^{2018} \abs{d_{x,t}-e_{x,t}\widehat{\mu}_{x t}},\quad x=50,\dots,90,
\end{equation*}
where $\widehat{\mu}_{x t}$ is the posterior mean of the forecasted death rates. \autoref{mae} shows the MAE by age for France, the UK, the USA and Japan according to the standard BMA, stacking and pseudo-BMA. Shifting from BMA to stacking or pseudo-BMA, a large improvement in the forecasts accuracy is obtained, especially for the ages $70$ to $90$. In particular, for France and USA, the MAE levels clearly for the stacking approach lie below the MAE levels of BMA and Pseudo-BMA. For the UK, the performances of the three methods are close for the ages $50$ to $80$ but the stacking approach leads to better MAEs after age $80$. For Japan, the comparison is not obvious. However, we did compute the overall MAE across ages and years and found for Japan:
\begin{equation*}
	\text{MAE (BMA)}=579.30, \quad \text{MAE (stacking)}=489.74, \quad \text{MAE (Pseudo-BMA)}=758.76
\end{equation*}
Hence, at the aggregate level, stacking still provides a better forecast performance than BMA, even for Japan. 

\textcolor{black}{Finally, to assess the accuracy of the predictive forecast distribution for future death rates, we study scoring rules as considered in \autoref{eq:logS} and \autoref{eq:CPRS} in the simulation study. \autoref{tablescoring} presents the log score and the CRPS for France, UK, USA and Japan for the three model averaging approaches and all single models. First, we observe that stacking outperforms BMA and pseudo-BMA for France and USA while for Japan, BMA is the best aggregation model. For UK, the result is not clear: stacking is better in terms of CRPS but not in terms of log score. Concerning single models, there is no evidence of a \textit{best} single model across countries and the `optimal' model depends on the country and the scoring rule studied. We also note that stacking does not outperform all single models but tends to consistently rank among the top three. In this sense, stacking allows to reduce partially the model risk. }

\begin{table}[h!]
	\centering
	\caption{Log score and CRPS for France, UK, USA and Japan via BMA, stacking and pseudo-BMA, and all single models averaged over forecast years and ages. We indicate in bolds the best performance by model averaging approach and by single model. CRPS are multiplied by a factor 1000 for clarity.} \label{tablescoring}
	\begin{tabular}{@{}llllll@{}}
		\toprule
		\multicolumn{1}{c}{\multirow{2}{*}{}} & \multicolumn{2}{c}{France}                               &                      & \multicolumn{2}{c}{UK}                                   \\ \cmidrule(lr){2-3} \cmidrule(l){5-6} 
		\multicolumn{1}{c}{}                  & \multicolumn{1}{c}{Log score} & \multicolumn{1}{c}{CRPS} & \multicolumn{1}{c}{} & \multicolumn{1}{c}{Log Score} & \multicolumn{1}{c}{CRPS} \\ \midrule
		BMA                                   & -4.585                        & 0.988                    &                      & \textbf{-5.275  }                      & 2.109                    \\
		Stacking                              & \textbf{-4.748}                        & \textbf{0.848 }                   &                      & -5.006                        & \textbf{1.965}                    \\
		Pseudo-BMA                            & -4.585                        & 0.988                    &                      & \textbf{-5.275  }                      & 2.109                    \\ \midrule
		LC                                    & -4.943                        & 1.050                    &                      & -4.701                        & 1.840                    \\
		RH                                    & -4.585                        & \textbf{0.988  }                  &                      & -4.743                        & 1.854                    \\
		APC                                   & \textbf{-5.402}                        & 1.441                    &                      & -5.019                        & 2.256                    \\
		CBD                                   & -4.183                        & 3.860                    &                      & -4.401                        & \textbf{1.539 }                   \\
		M6                                    & -3.799                        & 3.298                    &                      & \textbf{-5.275}                        & 2.109                    \\ \midrule
		\multirow{2}{*}{}                     & \multicolumn{2}{c}{USA}                                  &                      & \multicolumn{2}{c}{Japan}                                \\ \cmidrule(lr){2-3} \cmidrule(l){5-6} 
		& \multicolumn{1}{c}{Log score} & \multicolumn{1}{c}{CRPS} & \multicolumn{1}{c}{} & \multicolumn{1}{c}{Log Score} & \multicolumn{1}{c}{CRPS} \\ \midrule
		BMA                                   & -3.815                        & 1.793                    &                      & \textbf{-5.166  }                      & \textbf{1.084  }                  \\
		Stacking                              & \textbf{-3.956 }                       & \textbf{1.441 }                   &                      & -4.972                        & 1.430                    \\
		Pseudo-BMA                            & -3.804                        & 1.771                    &                      & -5.158                        & 2.035                    \\ \midrule
		LC                                    & -3.168                         & 4.166                    &                      &\textbf{ -5.276  }                      & 1.164                    \\
		RH                                    & -3.815                        & 1.793                    &                      & -5.166                        & \textbf{1.084 }                   \\
		APC                                   & \textbf{-5.551  }                      & \textbf{0.938  }                  &                      & -5.158                        & 2.035                    \\
		CBD                                   & -3.703                        & 2.165                    &                      & -4.972                        & 2.577                    \\
		M6                                    & -5.151                        & 1.404                    &                      & -4.539                        & 2.068                    \\ \bottomrule
	\end{tabular}
\end{table}

\textcolor{black}{Overall, this validation exercise shows that stacking tends to outperform Pseudo-BMA and standard BMA in terms of the ability to predict $10$-year ahead for the four countries considered here. We remark that for Japan, the situation is not evident: the MAE is better for stacking but the scoring rules give the best performance to the standard BMA. Moreover, the performance of standard BMA and Pseudo-BMA appears similar except that standard BMA performs better for Japanese mortality data. In summary, this section shows that a model which provided good forecasts for the last $10$ years has a good chance to perform well for the following $10$ years. On the other hand, a model that fits well the mortality data has no a priori reason to be good at \textit{forecasting} future mortality data. We therefore recommend stacking based on leave-future-out validation to methods based on goodness-of-fit (standard BMA) for forecasting purposes.}

\section{Impact of Covid-type effect on mortality forecasting}\label{sec:covid}

In the context of the recent Covid-19 pandemic, it is important to determine how mortality models and forecasts react to a pandemic shock. In the following, we have perturbed the French male data with two years of excess mortality followed by one year of lower mortality, and assessed the impact in terms of model averaging weights and life expectancy. This pandemic scenario is in the spirit of \cite{cairns2020impact} who proposed an accelerated deaths model to explore the impacts of the pandemic on life expectancy. The authors argue that ``\textit{many of those who die from coronavirus would have died anyway in the relatively near future due to their existing frailties or  co-morbidities}. \textit{Therefore, the life expectancy of the surviving population might slightly increase compared to their pre-pandemic levels}". For this reason, we do compensate two years of excess mortality by a slight decrease in mortality in the third year.

We take the male death data for France until year $2018$ from the Human Mortality Database, and perturb the death counts associated to the remaining three years as follows:
\begin{itemize}
	\item For the years 2016 and 2017, we assume that there is a uniform death increase of $5\%$ across all ages:
	\begin{equation*}
		d_{x,t}^{\text{new}}=(1+\beta)  d_{x,t},
	\end{equation*}
	with $\beta=0.05$ for $t=2016,2017$.
	
	\item The increase in deaths is then compensated with a year of lower mortality. We assume a death decrease of 2\% across ages: 
	\begin{equation*}
		d_{x,t}^{\text{new}}=(1-\beta)  d_{x,t},
	\end{equation*}
	with $\beta=0.02$ for $t=2018$.
\end{itemize}

First, we derive the weights associated to the standard BMA (marginal likelihood), stacking and pseudo-BMA approaches based on $40$ years of data (1979-2018) including $10$ validation years (2009-2018) with and without perturbations. Different observations can be drawn from the results in \autoref{perturbations}. For BMA and Pseudo-BMA, the perturbations do not affect the weights: the Renshaw-Haberman model is chosen by the BMA approach and the APC model is favored by the Pseudo-BMA approach. For the stacking approach, we observe some slight changes in the weights. With the perturbations, some weight is given to the Lee-Carter model and the stacking approach therefore averages over three models (LC, RH and APC). Moreover, we note that the weights obtained in \autoref{perturbations} are different from the ones obtained in \autoref{weightcountries} in the previous section since the validation and calibration periods are different. For instance, for Pseudo-BMA, RH was chosen for the validation period 1999-2008 while APC was the selected model for the validation period 2009-2018.

\begin{table}[h]
	\centering
	\caption{Model Weights for France with and without Covid-type effect.} \label{perturbations}
	\begin{tabular}{@{}ccccccc@{}}
		\toprule
		& \multicolumn{2}{c}{BMA} & \multicolumn{2}{c}{Stacking} & \multicolumn{2}{c}{Pseudo-BMA} \\ \midrule
		Perturbations & Without      & With     & Without        & With        & Without         & With         \\ \midrule
		LC            & 0            & 0        & 0              & 0.147       & 0               & 0            \\
		RH            & 1            & 1        & 0.21           & 0.170        & 0               & 0            \\
		APC           & 0            & 0        & 0.79           & 0.682       & 1               & 1            \\
		CBD           & 0            & 0        & 0              & 0           & 0               & 0            \\
		M6            & 0            & 0        & 0              & 0           & 0               & 0            \\ \bottomrule
	\end{tabular}
\end{table}

\begin{figure}[h!]
	\begin{center}
		\includegraphics[scale=0.75]{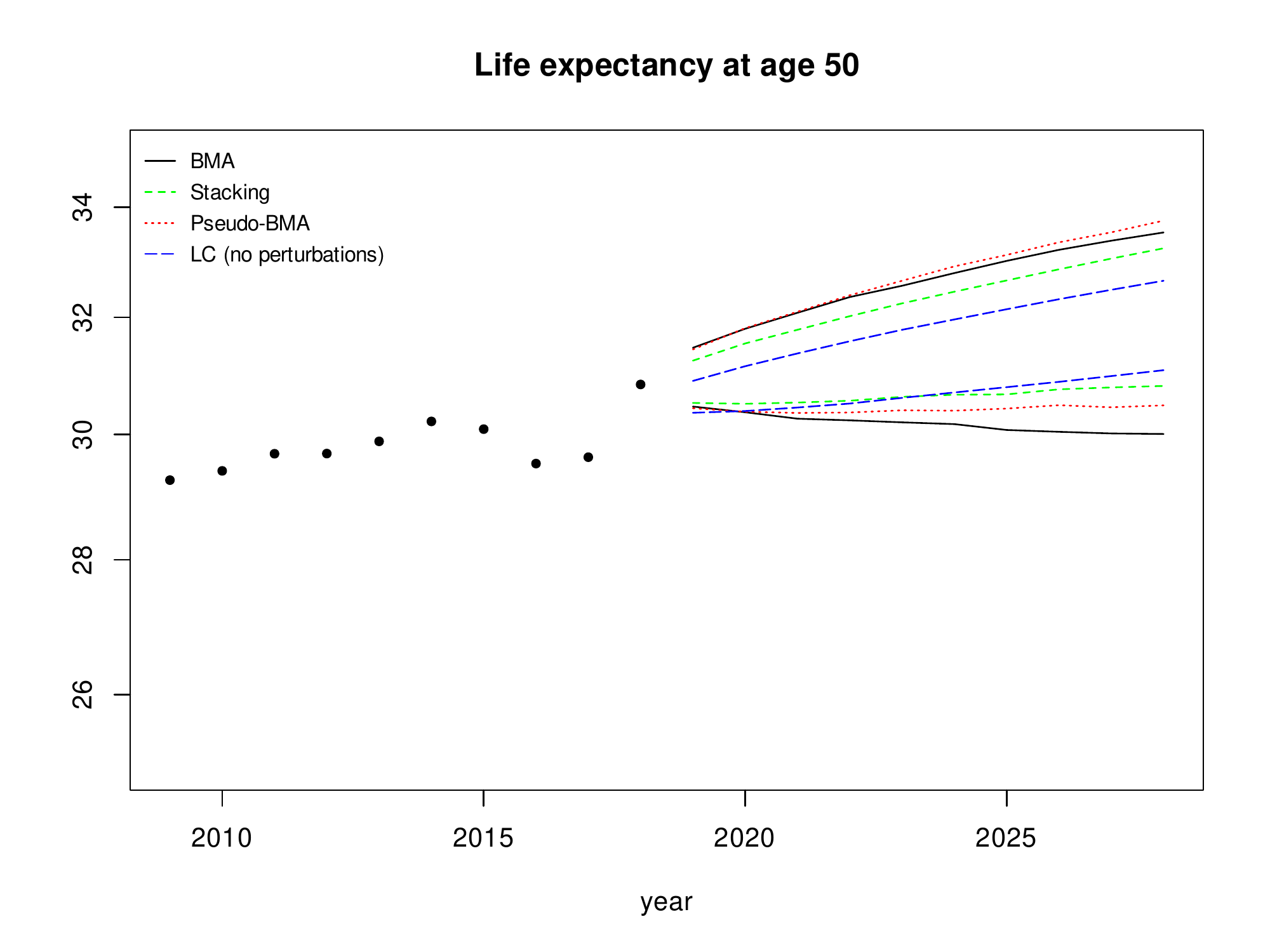}
		\caption{95\% prediction intervals for the life expectancy at age 50 (truncated at age 90) for the 10-year period 2019-2028 via the three model averaging approaches with perturbed data. For comparison, we also provide the 95\% prediction intervals via the Lee-Carter (LC) model without perturbations.}
		\label{perturbations3}
	\end{center}
\end{figure}

To measure the effect on life expectancy, and since we focus on the age range $50-90$, we compute the life expectancy at age $50$ truncated at age $90$ for the next $10$ years (2019-2028), that is
\begin{equation}
	e_{50:\actuarialangle{40},t}=\sum_{k=1}^{40} \px[k]{50,t} \label{truncatedlifeexpectancy}
\end{equation}
where $\px[k]{50,t}$ is the $k$-year survival probability at year $t$ just like in Equation \eqref{survivalprobability}. We note that \eqref{truncatedlifeexpectancy} can be interpreted as the average number of payments of a life annuity at age $50$ that ends at age $90$ since
\begin{equation}
	e_{50:\actuarialangle{40}}=\mathbb{E} \left[\min \left(K_{50},40\right)\right]
\end{equation}
where $K_{50}$ is the number of years lived by a person aged 50 (see for instance Section 2.6 in \cite{dickson2013actuarial}). In \autoref{perturbations3}, we plot the life expectancies, observed and predicted, from 2009 to 2018, according to each model averaging method. In order to better assess the impact of the perturbations on the overall uncertainty, we show the predictions of the Lee-Carter model \textit{without} perturbations.  In particular, we observe that the perturbed data via all three approaches produce larger confidence intervals compared to the baseline LC model without perturbations as one would expect. Indeed, the perturbations increase the volatility of the period effects $\bm{\kappa}_{t}$ and therefore the uncertainty in future life expectancy. 

\textcolor{black}{The median life expectancy with and without Covid-type effect is plotted on \autoref{perturbations2}. With the perturbations, the median life expectancy increases between the years 2019 and 2028, and this increase is more important than the situation without Covid-type effect. Hence, we do observe a compensation effect of the pandemic. }

\begin{table}[h]
	\centering
	\caption{Median Life Expectancy at age 50 (truncated at age 90) for French Male with and without Covid-type effect.} \label{perturbations2}
\begin{tabular}{@{}ccccccc@{}}
	\toprule
	& \multicolumn{2}{c}{BMA} & \multicolumn{2}{c}{Stacking} & \multicolumn{2}{c}{Pseudo-BMA} \\ \midrule
	Perturbations & Without    & With       & Without       & With         & Without        & With          \\ \midrule
	2019          & 30.59                       & 30.97                    & 30.57                       & 30.89                    & 30.57                       & 30.95                    \\
	2020          & 30.70                       & 31.11                    & 30.71                       & 31.04                    & 30.72                       & 31.10                    \\
	2021          & 30.80                       & 31.23                    & 30.84                       & 31.19                    & 30.86                       & 31.26                    \\
	2022          & 30.89                       & 31.33                    & 30.98                       & 31.33                    & 31.00                       & 31.42                    \\
	2023          & 30.97                       & 31.43                    & 31.10                       & 31.47                    & 31.12                       & 31.55                    \\
	2024          & 31.05                       & 31.52                    & 31.22                       & 31.60                    & 31.27                       & 31.71                    \\
	2025          & 31.10                       & 31.61                    & 31.32                       & 31.73                    & 31.39                       & 31.84                    \\
	2026          & 31.19                       & 31.70                    & 31.43                       & 31.86                    & 31.49                       & 31.99                    \\
	2027          & 31.22                       & 31.80                    & 31.52                       & 31.99                    & 31.62                       & 32.11                    \\
	2028          & 31.30                       & 31.91                    & 31.62                       & 32.10                    & 31.73                       & 32.23                    \\ \bottomrule
\end{tabular}
\end{table}

\begin{figure}[h]
	\centering
	\includegraphics[width=0.8\textwidth]{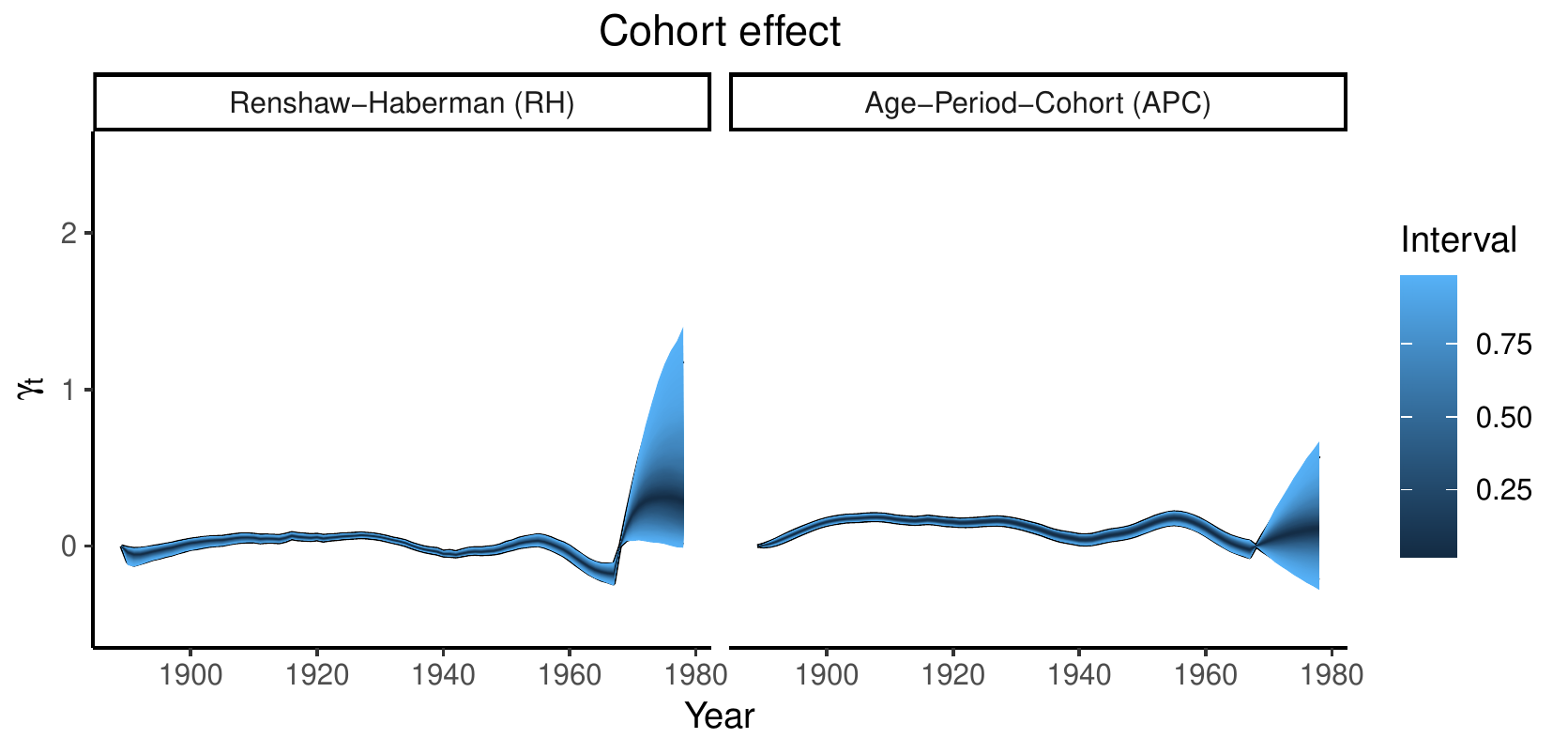}
	\caption{95\% prediction intervals for the cohort parameter $\gamma_{t}$ in the RH and APC models.\label{perturbations5}}
\end{figure}

\textcolor{black}{Overall, we find that the three model averaging approaches predict an increase in life expectancy which is consistent with the historical trend and the $5\%$ decrease in the number of deaths associated to the \textit{compensation effect} of the pandemic. We also note that the BMA approach which selects, in this case, the RH model provides wider prediction intervals due to the cohort effect as depicted in \autoref{perturbations5}.}

\section{Conclusion}\label{sec:conclusion}
In this work, we address the problem of stochastic mortality model averaging. We start by setting up an attractive Bayesian modeling framework because it allows us to consider several mortality models and to account for the uncertainty around the parameter estimates. Model averaging strategies are then applied to mitigate the risk of selecting the wrong model. The standard Bayesian model averaging, based on how well the model fits the training dataset is challenged by two other model averaging strategies, referred to as stacking and pseudo-BMA, that focus on the out-of-sample error.\\

\noindent We recommend the use of the leave-future-out based model averaging approaches for the purpose of forecasting mortality trends. Our study draws on extensive simulation study and applications to real-world mortality data sets (with and without COVID-like disruption).\\ 
 
\noindent This work could be extended in many interesting ways. First, the validation technique could be adapted to the case where the mortality patterns exhibit a change of regime. In fact, as discussed with the COVID-type impact, the model averaging approach should assign more weights to models that are not only good at representing the past but also at forecasting the future. Here, we should introduce some potential regime switching techniques into the considered models in order to tackle such a problem. However, this interesting problem is beyond the scope of the current paper and will be investigated in a future work. Finally, given the ability of the averaging techniques to accommodate classic and most used models, an R package implementing the three model averaging approaches is available for download to researchers as well as practitioners at \url{https://CRAN.R-project.org/package=StanMoMo}.

\section*{Acknowledgment}
The authors would like to thank the Editor and two anonymous referees who
provided useful and detailed comments that substantially improved the current manuscript. This work was supported by the Joint Research Initiative on   ``Mortality Modeling and Surveillance” funded by AXA Research Fund. S. Loisel and Y. Salhi also acknowledge support from the BNP Paribas Cardif Chair “New Insurees, Next Actuaries” (NINA) and the Milliman research initiative ``Actuariat Durable”. Y. Salhi benefited from the support of the CY Initiative of Excellence (grant “Investissements d’Avenir” ANR-16-IDEX-0008), Project “EcoDep” PSI-AAP2020--0000000013. P-O. Goffard's work is partially funded by the DIALog – Digital Insurance And Long-term risks – Chair under the aegis of the Fondation du Risque, a joint initiative by UCBL and CNP Assurances.

\end{document}